%% file: cases_tecs19.tex
\documentclass[sigconf]{acmart}

\usepackage{xcolor}
\usepackage[linesnumbered,ruled]{algorithm2e}

\SetCommentSty{mycommfont}
\usepackage{multirow}
\usepackage{xcolor}
\usepackage{varwidth}
\usepackage{pifont}
\let\oldding\ding
\renewcommand{\ding}[2][1]{\scalebox{#1}{\oldding{#2}}}
\newtheorem{Tlemma}{\small \underline{Policy}}

\newcommand{\ineq}[1]{\footnotesize$#1$\normalsize}{}
\newcommand{\tech}{\text{{PALP}}}{}
\newcommand{\mpat}{\text{{MultiPartition}}}{}
\newcommand{\alat}{\text{{\ineq{\color{blue} 23\%}}}}{}
\newcommand{\perf}{\text{{\ineq{\color{blue} 28\%}}}}{}
{}

\newcommand{\hl}[1]{\textcolor{black}{#1}}
\newcommand{\mc}[1]{\textcolor{black}{#1}}
\newcommand{\fmc}[1]{\textcolor{black}{#1}}

\AtBeginDocument{%
  \providecommand\BibTeX{{%
    \normalfont B\kern-0.5em{\scshape i\kern-0.25em b}\kern-0.8em\TeX}}}

\setcopyright{acmcopyright}
\copyrightyear{2019}
\acmYear{2019}
\acmDOI{10.1145/1122445.1122456}

\acmJournal{TWEB}
\acmVolume{9}
\acmNumber{4}
\acmArticle{39}
\acmYear{2010}
\acmMonth{3}
\copyrightyear{2009}

\setcopyright{acmlicensed}

\acmDOI{0000001.0000001}

\setcopyright{none}
\makeatletter
\def\@copyrightspace{\relax}
\makeatother
\renewcommand\footnotetextcopyrightpermission[1]{} 
\acmConference[]{}{}{}
\acmBooktitle{}

\begin{document}

\title{Enabling and Exploiting Partition-Level Parallelism (PALP) in Phase Change Memories}

\author{Shihao Song}
\email{shihao.song@drexel.edu}
\author{Anup Das}
\orcid{0000-0002-5673-2636}
\email{anup.das@drexel.edu}
\affiliation{%
  \institution{Drexel University}
  \streetaddress{3120 Market Street}
  \city{Philadelphia}
  \state{Pennsylvania}
  \postcode{19104}
}
\author{Onur Mutlu}
\email{omutlu@gmail.com}
\affiliation{%
  \institution{ETH Z{\"u}rich}
  \city{Z{\"u}rich}
  \state{Switzerland}
  \postcode{43017-6221}
}
\author{Nagarajan Kandasamy}
\email{kandasamy@drexel.edu}
\affiliation{%
  \institution{Drexel University}
  \streetaddress{3120 Market Street}
  \city{Philadelphia}
  \state{Pennsylvania}
  \postcode{19104}
}

\thanks{This article appears as part of the ESWEEK-TECS special issue and was presented in the International Conference on Compilers, Architecture, and Synthesis for Embedded Systems (CASES) 2019}








\renewcommand{\shortauthors}{Song et al.}

\begin{abstract}
 \input{sections/abstract.tex}
\end{abstract}

\ccsdesc[500]{Computer systems organization~Architectures}
\ccsdesc[500]{Hardware~Memory and dense storage}

\keywords{Phase-change memories (PCM), Sense Amplifiers, Write Drivers}

\maketitle


\section{Introduction}
\label{sec:introduction}
\input{sections/introduction}

\section{Background on PCM}
\label{sec:background}
\input{sections/background}

\section{Enabling Partition-Level Parallelism in PCM}
\label{sec:enable_PALP}
\input{sections/enable_palp}

\section{Exploiting Partition-Level Parallelism in PCM}
\label{sec:exploit_PALP}
\input{sections/exploit_palp}

\section{Evaluation Methodology}
\label{sec:evaluations}
\input{sections/evaluation_methodology}

\section{Results and Discussion}
\label{sec:results}
\input{sections/results}


\section{Related Works}
\label{sec:related_works}
\input{sections/related_works}

\section{Conclusion}
\label{sec:conclusions}
\input{sections/conclusions}

\bibliographystyle{ACM-Reference-Format}
\bibliography{pcm}










\end{document}

%% file: sections/abstract.tex
Phase-change memory (PCM) devices have multiple banks to serve memory requests \emph{in parallel}. Unfortunately, if two requests go to the \emph{same bank},  they have to be served \emph{one after another}, leading to lower \emph{system performance}.
We observe that a modern PCM bank is implemented as a collection of \emph{partitions} that operate mostly \emph{independently} while \emph{sharing} a \emph{few} global peripheral structures, which include the sense amplifiers (to read) and the write drivers (to write). 
\fmc{Based on this observation, we propose \textbf{\tech{}}, a \emph{new} mechanism that \textit{enables} partition-level parallelism within each PCM bank, and \emph{exploits} such parallelism 
by using the memory controller's access scheduling decisions. \tech{} consists of  \emph{three} new contributions.} \textbf{First}, 
we introduce \emph{new} PCM commands to 
enable parallelism in a bank's partitions in order to 
resolve the \textbf{read-write} bank conflicts, with \textit{no} changes needed to PCM logic or its interface. 
\textbf{Second}, we propose \textit{simple} circuit modifications that
introduce a new operating mode for the write drivers, in addition to their \textit{default} mode of serving write requests. When configured in this new mode, the write drivers can 
resolve the \textbf{read-read} bank conflicts,
working \textit{jointly} with the sense amplifiers.
\textbf{Finally}, we propose a \emph{new} access scheduling mechanism in PCM that improves performance by prioritizing those requests that exploit {partition-level parallelism} over other requests, including the long {outstanding} ones. 
While doing so, the memory controller also guarantees starvation-freedom and the PCM's running-average-power-limit (RAPL).

We evaluate \tech{} with workloads from the MiBench and SPEC CPU2017 Benchmark suites.
Our results show that \tech{} reduces average PCM access latency by \alat{}, and improves average system performance by \perf{} compared to the state-of-the-art approaches. 

%% file: sections/introduction.tex
Modern phase change memory (PCM) devices \cite{fagot2018low,lung2016double,CastellaniIMW16,navarro2018phase,villa2018pcm,LeeISCA2009} 
can serve multiple requests in parallel using different PCM banks \cite{lung2016double}. 
Unfortunately, when {two} requests go to the \emph{same bank}, they have to be served \emph{serially}. This is known as \emph{bank conflict}. Bank conflicts reduce \emph{system performance} by lowering the PCM \emph{bandwidth utilization}, causing CPU cores to \emph{stall}. To estimate the impact of bank conflicts on performance, we plot the distribution of \emph{read-read}, \emph{read-write}, and \emph{write-write} bank conflicts in 15 8-core workloads for a PCM memory of 8GB capacity and with eight 1GB banks (see our evaluation methodology in Section \ref{sec:evaluations}). We make the following two observations.

\begin{figure}[h!]
	\centering
	\centerline{\includegraphics[width=0.99\columnwidth]{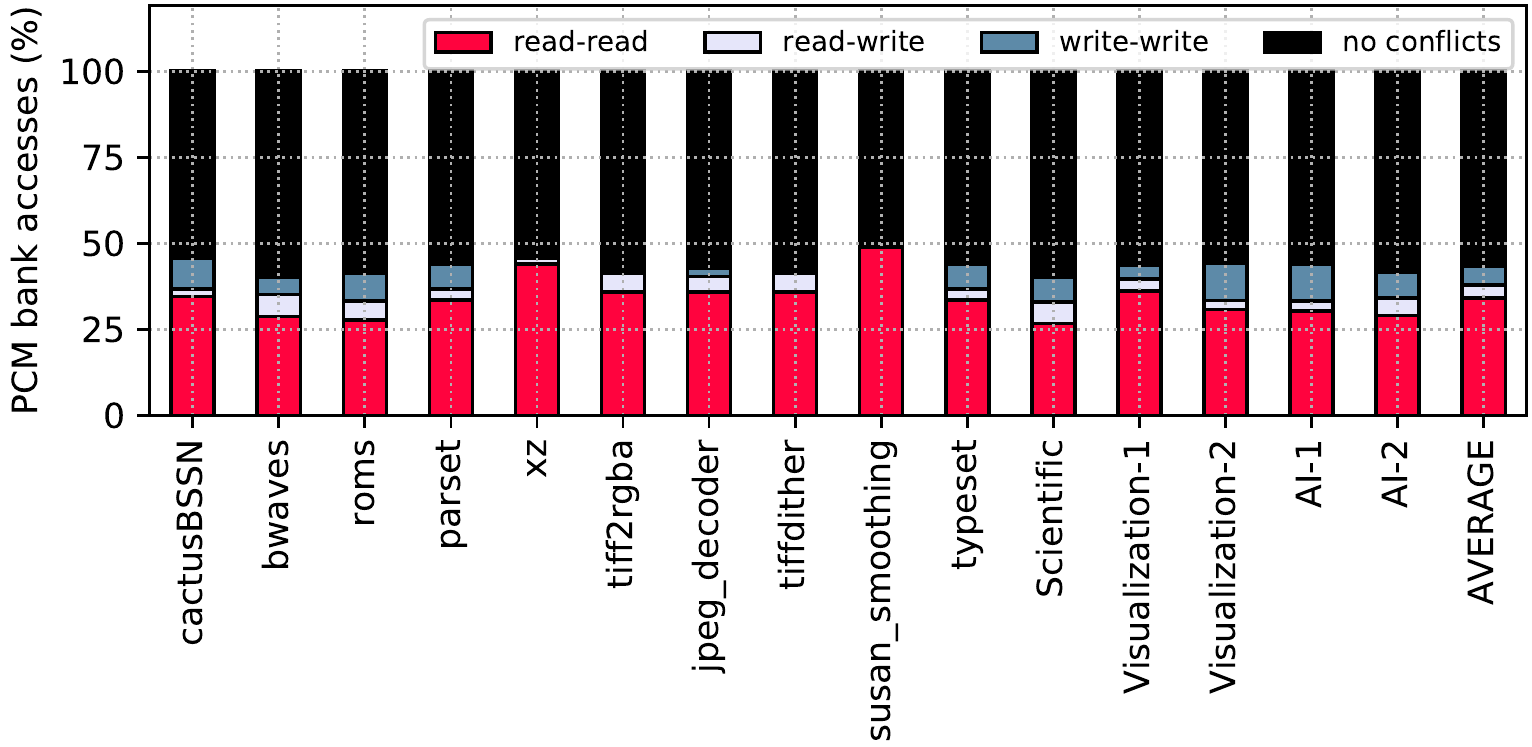}}
	\vspace{-10pt}
	\caption{Distribution of PCM bank conflicts for our 15 evaluated 8-core workloads (see Table \ref{tab:spec}).}
	\label{fig:motivation_plp}
\end{figure}

\hl{
	\textbf{First,} on average, 43\% of PCM requests in these workloads generate bank conflicts. This is due to the high temporal and spatial \textit{access locality} in these workloads that lead to repeated accesss to multiple rows that map to the same bank. 
	\textbf{Second}, read-read bank conflicts \textit{outnumber} read-write and write-write bank conflicts for all workloads, averaging to 34\% of all PCM requests (79\% of all bank conflicts). The read-write and write-write bank conflicts are fewer than read-read bank conflicts because write requests are fewer to begin with as they go through an extra level of cache implemented as eDRAM. The read requests, on the other hand, go directly to PCM, bypassing the eDRAM cache.
	We \textbf{conclude} that by \textit{resolving} both read-read and read-write bank conflicts in PCM, system performance can be improved significantly. We demonstrate \perf{} average performance improvement in Section \ref{sec:performance_improvement} for our evaluated workloads. As core count increases, bank conflicts also increase and become a bigger performance bottleneck \cite{hassan2016chargecache,KimISCA12,hassan2019crow}.
}

Our \textbf{goal} is to improve performance by resolving bank conflicts in PCM devices. To this end, we analyze the internal architecture of a PCM bank (see Figure \ref{fig:peripheral} for details).
\hl{We observe that a PCM bank is implemented as a collection of \emph{partitions} \cite{lung2016double} that operate mostly \emph{independently} while \emph{sharing} a few global peripheral structures, which include the sense amplifiers (to read) and the write drivers (to write). A peripheral structure has high area overhead \fmc{primarily due to the} write driver logic \cite{cho2005programming,sandhu2018memory,dray2018high}, which needs to generate high current profiles for the SET (100\ineq{\mu}A) and RESET (300\ineq{\mu}A) programming of the PCM cells. Due to this \emph{high area overhead}, PCM manufacturers integrate only a small number of these structures per bank to improve PCM density \cite{villa2018pcm}. For instance, in the recent PCM prototypes \cite{lung2016double,villa2018pcm}, there are only 128 peripheral structures per bank, \fmc{which are shared across all 16 partitions in the bank}.} 

Based on our key observation, we propose \textbf{\tech{}}, a \textit{new} mechanism that \emph{enables} partition-level parallelism in a PCM bank, and \emph{exploits} such parallelism by modifying \emph{only} the memory controller's access scheduling decisions.
We make the following three novel contributions.

\noindent\textbf{Contribution 1:} We propose a \emph{new} mechanism to {enable} read-write parallelism in a PCM bank's partition using the bank's peripheral structures to resolve the \emph{read-write bank conflicts}.
To this end, we introduce a \textit{new} PCM command called \texttt{READ-WITH-WRITE (RWW)}, and provide its detailed timing requirements.

\noindent\textbf{Contribution 2:} We propose \textit{simple} circuit modifications to introduce a new operating mode for the write drivers, with which they can also serve read requests. When configured in this new mode, the write drivers can resolve the \emph{read-read bank conflicts}, working \emph{jointly} with the sense amplifiers. To this end, we introduce a \textit{new} PCM command called \texttt{READ-WITH-READ (RWR)}, and provide its detailed timing requirements.

\noindent\textbf{Contribution 3:} We propose a \textit{new} memory access scheduling mechanism that prioritizes those PCM requests that exploit a PCM bank's partition-level parallelism, over other requests, including the {long} outstanding ones. This \textit{straightforward} and \textit{greedy} performance-oriented policy is controlled in \textit{two} ways. \textbf{First}, the memory controller ensures that {no} request is starved, i.e., backlogged {excessively}, forcefully serving the outstanding request otherwise, to guarantee starvation-freedom \cite{kim2010atlas}. \textbf{Second}, the memory controller ensures that the power consumption of the active partitions within the bank is {not} too high, forcefully serializing requests to the bank otherwise, to guarantee the running average power limit (RAPL) \cite{david2010rapl}.

We implement \tech{} for DRAM-PCM hybrid main memory systems \cite{QureshiISCA09,dhiman2009pdram,liu2018crash,li2017utility,zhao2014firm,yoon2012row,meza2012enabling,ham2013disintegrated,bock2016concurrent,lin2017hybrid,yu2017banshee,meza2013case}, which use DRAM as a cache to PCM. 
Given the still \emph{speculative} state of the PCM technology, we describe \tech{} based on the architecture and memory timings of IBM's \ineq{20nm} PCM prototype \cite{lung2016double}.
We evaluate \tech{} with workloads from the MiBench \cite{guthaus2001mibench} and SPEC CPU2017 \cite{bucek2018spec} benchmark suites. Our results show that \tech{} reduces average PCM access latency by \alat{}, and improves average system performance by \perf{} compared to the state-of-the-art approaches.
As \tech{} exploits the PCM bank's partition-level parallelism using the memory controller's access scheduling decision, it can be easily combined with orthogonal mechanisms: (1) those aiming to reduce the \emph{number} of write accesses to PCM \cite{dhiman2009pdram,li2017utility,bock2016concurrent}, and (2) those aiming to \emph{improve} PCM's cell endurance \cite{SeongSecurityISCA2010,akram2018write}.

{Although our work is inspired by the notion of subarray-level parallelism (SALP) in DRAM \cite{KimISCA12}, exploiting parallelism within PCM banks is \textit{unique} in the following \emph{two} aspects. \textbf{First}, while \emph{many} subarrays can be active simultaneously in a DRAM, PCM's peripheral structures allow only \emph{two} partitions to be active simultaneously in a PCM bank \cite{barkley2017apparatus}. \textbf{Second}, while subarray-level parallelism in DRAM can resolve \textit{any} bank conflict, partition-level parallelism in PCM can resolve only the \textit{read-read} or \textit{read-write} bank conflicts. These unique properties of \tech{} lead to different performance trade-offs, which we present in Section \ref{sec:results}. 
}

The closest state-of-the-art PCM  mechanisms, such as \cite{zhou2016efficient,yue2013exploiting}, have \emph{only} addressed the read-write bank conflicts in PCM, assuming a simple \textit{first-come-first-serve} (FCFS) scheduling policy \cite{RixnerISCA2000}. We demonstrate in Section \ref{sec:fcfs_limitations} that the FCFS policy \textit{cannot} {efficiently} exploit each PCM bank's partition-level parallelism.
We not only resolve \textit{both} the read-read and read-write bank conflicts in PCM, but also develop a \textit{new} performance-oriented scheduling policy to \textit{better} exploit such partition-level parallelism. \hl{Our \tech{} design is based on IBM's PCM chip interfaced with ARMv8-A (aarch64) processor using the DDR4 protocol.
} In Table \ref{tab:compare_sota}, we summarize the state-of-the-art mechanisms and highlight the contributions of this paper.

\begin{table}[h!]
	\renewcommand{\arraystretch}{1.2}
	\setlength{\tabcolsep}{2pt}
	\centering
	{\fontsize{6}{10}\selectfont
	\begin{tabular}{|c|c|c|c|c|c|c|}
	\hline
	\multirow{2}{*}{\textbf{Mechanism}} & \textbf{Read-Write} & \textbf{Read-Read} & \multirow{2}{*}{\textbf{Scheduling}} & \textbf{Memory} & \textbf{Memory} & \textbf{Memory}\\
	& \textbf{Bank Conflicts} & \textbf{Bank Conflicts} & & \textbf{Interface} & \textbf{Technology} & \textbf{Power}\\
	\hline
	\hline
	Baseline \cite{ArjomandISCA16} & $\times$ & $\times$ & FCFS & DDR2 & PCM & not controlled\\
	\cite{zhou2016efficient,yue2013exploiting} & $\surd$ & $\times$ & FCFS & not considered & PCM & not controlled\\
	SALP \cite{KimISCA12} &  $\surd$ & $\surd$ & FR-FCFS & DDRx & DRAM & not controlled\\
	\hline
	\tech{} & 	$\surd$ & 	$\surd$ & new & DDRx & 	PCM & RAPL\\
	\hline
	\end{tabular}}
	\caption{Comparison of \tech{} with state-of-the-art PCM and DRAM mechanisms.}
	\label{tab:compare_sota}
\end{table}



%% file: sections/background.tex
This section provides a brief background on PCM organization and operation required to understand \tech{}. PCM, like DRAM, is organized \emph{hierarchically} \cite{lung2016double}. An example PCM memory of \ineq{128GB} capacity has \ineq{4} {channels}, with 4 ranks per channel, and 8 banks per rank. 
A PCM bank has \ineq{8} {partitions}, each of which is an arrays of \ineq{4096} wordlines and \ineq{256K} bitlines.
A PCM bank has \ineq{128} peripheral structures, which include the sense amplifiers (to read) and the write drivers (to write). The peripheral structures in PCM banks allow to read and program 128 PCM cells \emph{in parallel}.
Therefore, the read and write granularity is \ineq{128} bits (the size of a \emph{memory line}). 

A PCM cell is built with the chalcogenide alloy (e.g., Ge${}_2$Sb${}_2$Te${}_5$ \cite{ovshinsky1968reversible}), and is connected to a bitline and a wordline using an access device. The amorphous phase (RESET) in this alloy has higher resistance than the crystalline phase (SET).
To RESET a PCM cell, a high current pulse of \emph{short duration} is applied and \emph{quickly terminated}. 
To SET a PCM cell, the chalcogenide alloy is heated above its crystallization temperature, but below its melting point for a \emph{long duration}. 
Finally, to read the content of a PCM cell, a \emph{small} electrical pulse is applied \textit{without} inducing any phase change in the material. To serve a memory request that accesses data at a particular row and column address within a partition, a memory controller issues \emph{three} commands to a PCM bank.
\begin{itemize}
	\item \texttt{\underline{ACTIVATE}(A):} activate the wordline and enable the access device for the PCM cells to be accessed.
	\item \texttt{\underline{READ}(R)/\underline{WRITE}(W):} drive read or write current through the PCM cell. After this command executes, the data stored in the PCM cell is available at the output terminal of the sense amplifier, or the write data is programmed to the PCM cell.
	\item \texttt{\underline{PRECHARGE}(P):} deactivate the wordline and bitline, and prepare the bank for the next access. 
\end{itemize}
The \texttt{A}-\texttt{A} interval (tRC) for the same bank is \ineq{47} cycles for a write request, and \ineq{19} cycles for a read request; \texttt{A}-\texttt{W/R} (tRCD) is 1 cycle; the read latency (RL) is 10 cycles; the write latency (WL) is 3 cycles; and the write recovery time (tWR) is 35 cycles. These timing parameters are based on a {266MHz} memory clock and a DDR2 interface \cite{lung2016double} (see also Table \ref{tab:simulation_parameters} for our simulation parameters).

%% file: sections/enable_palp.tex
\begin{figure}[h!]
	\centering
	\centerline{\includegraphics[width=0.99\columnwidth]{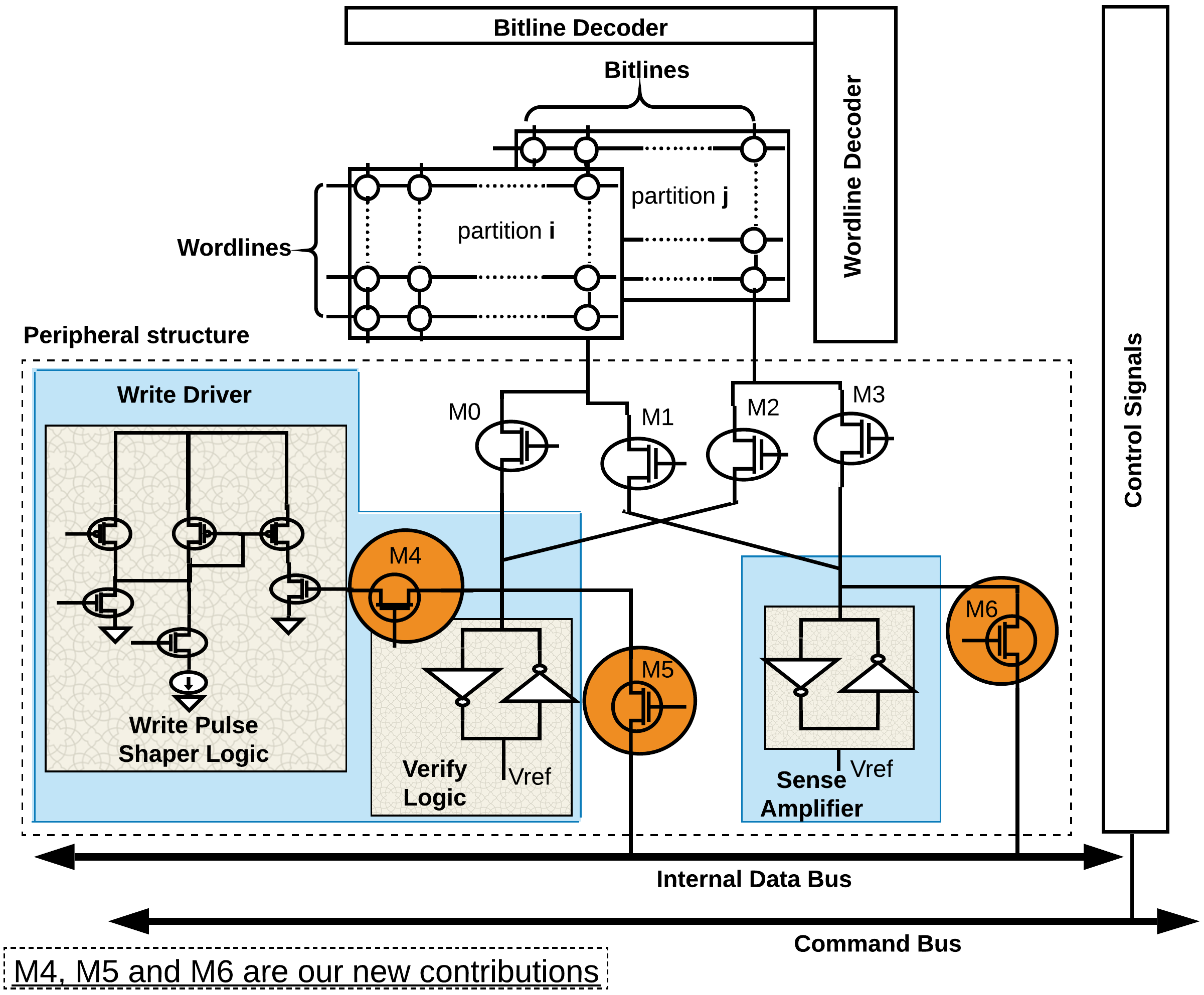}}
	\caption{Architecture of the new peripheral structure, which include a write driver and a sense amplifier. }
	\label{fig:peripheral}
\end{figure}

Figure \ref{fig:peripheral} illustrates a PCM bank's peripheral structures in detail \cite{fagot2018low,lung2016double,navarro2018phase,villa2018pcm,villa201045nm,goda2018programming}. 
There are 128 peripheral structures, which are shared across all bitlines in the bank. Each peripheral structure is connected to all the partitions in the bank.
To simplify our discussion, we illustrate only one peripheral structure, which is connected to the two partitions \ineq{i} and \ineq{j}. 
The peripheral structure contains the sense amplifier (to read) and the write driver (to write), which are connected to the two partitions using the NMOS transistors \texttt{M0, M1, M2,} and \texttt{M3}. 
These transistors are turned ON or OFF, based on the partition that needs to be accessed. The bitline and wordline decoders are used to connect the PCM cell at a particular row and column address to the peripheral structure.


\subsection{Resolving read-write bank conflicts in PCM}
\label{sec:rw}
Table \ref{tab:mos_control} reports how the transistors \texttt{M0, M1, M2,} and \texttt{M3} are configured to serve read and write requests.
Observe that only \emph{one} transistor is \texttt{ON} in the baseline PCM design (e.g., \cite{ArjomandISCA16}) to serve a read or write request from the bank.

\begin{table}[!h]
	\renewcommand{\arraystretch}{1.2}
	\centering
	{\fontsize{8}{10}\selectfont
	\begin{tabular}{|l||c|c|c|c|}
		\hline
		\textbf{Operation} & \textbf{\texttt{M0}} & \textbf{\texttt{M1}} & \textbf{\texttt{M2}} & \textbf{\texttt{M3}}\\
		\hline
		\hline
		write to partition i & \textcolor{blue}{ON} & OFF & OFF & OFF \\
		read from partition i & OFF & \textcolor{blue}{ON} & OFF & OFF \\
		write to partition j & OFF & OFF & \textcolor{blue}{ON} & OFF \\
		read from partition j & OFF & OFF & OFF & \textcolor{blue}{ON} \\
		\hline
	\end{tabular}}
\caption{Transistor configurations in the Baseline design for reading from or writing to a single partition.}
\label{tab:mos_control}
\end{table}

To investigate what is actually needed to resolve the read-write bank conflicts in PCM, we take a closer look at the connections of \texttt{M0, M1, M2,} and \texttt{M3}. Observe that if transistors \texttt{M0} and \texttt{M3} (or \texttt{M1} and \texttt{M2}) are simultaneously enabled, the sense amplifier can read a PCM cell from partition \ineq{j}, while the write driver is programming a PCM cell in partition \ineq{i} (or vice versa). 
Table \ref{tab:rw} summarizes our findings. Some transistor configurations can lead to data corruption. For example, if \texttt{M0 = OFF, M1 = ON, M2 = OFF, M3 = ON}, (the last entry in the table), then the data from two different partitions (i.e., two different PCM cells) will be connected to the sense amplifier. This compromises the integrity of the data to be read at the output of the sense amplifier. We mark all such table entries as \textit{invalid}. 

\begin{table}[!h]
	\renewcommand{\arraystretch}{1.2}
	\centering
	{\fontsize{8}{10}\selectfont
	\begin{tabular}{|l||c|c|c|c|}
		\hline
		\textbf{Operation} & \textbf{\texttt{M0}} & \textbf{\texttt{M1}} & \textbf{\texttt{M2}} & \textbf{\texttt{M3}}\\
		\hline
		\hline
		write to partition i, read from partition j & \textcolor{blue}{ON} & OFF & OFF & \textcolor{blue}{ON}\\
		read from partition i, write to partition j & OFF & \textcolor{blue}{ON} & \textcolor{blue}{ON} & OFF\\
		invalid & \textcolor{blue}{ON} & \textcolor{blue}{ON} & OFF & OFF\\
		invalid & OFF & OFF & \textcolor{blue}{ON} & \textcolor{blue}{ON}\\
		invalid & \textcolor{blue}{ON} & OFF & \textcolor{blue}{ON} & OFF\\
		invalid & OFF & \textcolor{blue}{ON} & OFF & \textcolor{blue}{ON}\\
		\hline
	\end{tabular}}
\caption{Transistor configurations to resolve the read-write bank conflicts in PCM.}
\label{tab:rw}
\end{table}


\hl{This is \textit{precisely} the parallelism we seek in PCM banks, which can resolve \textbf{read-write} bank conflicts, only when the two conflicting requests are to two different partitions within the bank.}
To this end, we propose the following two \emph{architectural} enhancements.
\begin{itemize}
	\item \textbf{First}, the memory controller must issue two back-to-back \texttt{ACTIVATE} commands to the PCM bank to \textit{decode} the wordline and bitline addresses in the two partitions. 
	\item \textbf{Second}, the memory controller must configure the transistors to connect the two partitions, one to the write drivers, and the other to the sense amplifiers.
\end{itemize}
To accomplish these actions, we introduce a \emph{new} PCM command:
\begin{itemize}
	\item \texttt{\underline{READ-WITH-WRITE} (RWW):} connect the PCM bank's sense amplifiers and write drivers to the two decoded partitions.
\end{itemize}

\begin{figure}[h!]
	\centering
	\centerline{\includegraphics[width=0.79\columnwidth]{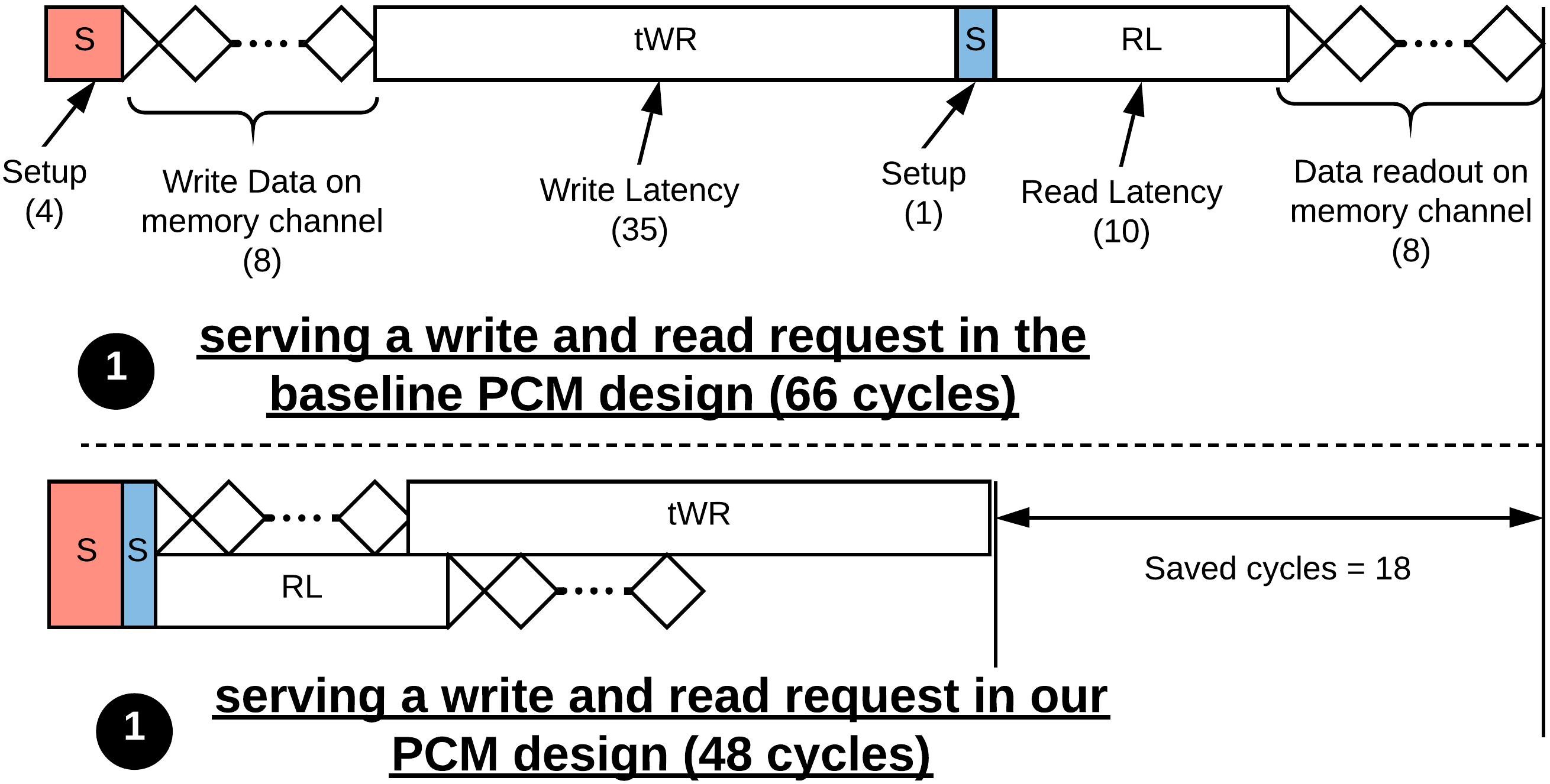}}
	\caption{Performance in the baseline (\ding{182}) and our PCM design (\ding{183}) when scheduling a write and a read request to different partitions in the same bank.}
	\label{fig:savings_rw}
\end{figure}

Figure \ref{fig:savings_rw} compares how a write and read request is scheduled in baseline PCM (\ding{182}) with our proposed PCM (\ding{183}) in which the memory controller exploits the PCM bank's partition-level parallelism. Following are the respective command sequences.

\begin{varwidth}[t]{0.5\textwidth}
	\footnotesize{Baseline PCM:}
	\begin{itemize}
		\item \texttt{ACTIVATE address in \ineq{i}}
		\item \texttt{WRITE}
		\item \texttt{PRECHARGE}
		\item \texttt{ACTIVATE address in \ineq{j}}
		\item \texttt{READ}
		\item \texttt{PRECHARGE}
		\item[--] Service time = 47 + 19 
		\item[]= 66 cycles
		\item[ ]
	\end{itemize}
\end{varwidth}
\hspace{2em}
\begin{varwidth}[t]{.5\textwidth}
	\footnotesize{Proposed PCM:}
	\begin{itemize}
		\item \texttt{ACTIVATE address in \ineq{i}}
		\item \texttt{ACTIVATE address in \ineq{j}}
		\item \texttt{READ-WITH-WRITE}
		\item \texttt{PRECHARGE}
		\item[--] Service time = 1 + 47 
		\item[ ] = 48 cycles
	\end{itemize}
\end{varwidth}

In the baseline PCM design, {\texttt{A}-\texttt{W}-\texttt{P}} takes 47 cycles and {\texttt{A}-\texttt{R}-\texttt{P}} takes 19 cycles (see Figure \ref{fig:savings_rw}), contributing to a total of 66 cycles to serve these two requests. 
In our PCM design, {\texttt{A}-\texttt{RWW}-\texttt{P}} takes 47 cycles, during which the write recovery time (tWR) partially overlaps with the read latency (RL). The extra 1 cycle is due to the second \texttt{ACTIVATE} command. The total service latency is 48 cycles for serving a read and write request mostly in parallel, a reduction of 18 cycles (27\%) over the baseline.

\subsection{Resolving read-read bank conflicts in PCM}
\label{sec:rr}
From the write driver's internal circuit diagram shown in Figure \ref{fig:peripheral}, we observe that the write driver can be viewed as a collection of two components -- the \textbf{write pulse shaper logic}, which generates the current pulses necessary for the PCM cell's SET and RESET operations, and the \textbf{verify logic}, which verifies the correctness of these operations. These two circuit components together serve write requests from the bank using a PCM write scheme known as \emph{program-and-verify}  (P\&V)~\cite{nirschl2007write,goda2018programming}. 
The verify logic essentially consists of two cross-coupled inverters, which can be configured as a sense amplifier, similar to the one that is already part of the peripheral circuit.
Based on this observation, we propose \emph{simple} circuit modifications to introduce the decoupling transistor \texttt{M4} (see Fig. \ref{fig:peripheral}), which can be configured when needed, to \textit{decouple} the verify logic from the write pulse shaper logic. As a result of this modification, we introduce \textit{two} operating modes for the write driver. In the \textit{decoupled mode}, the verify logic can serve read requests concurrently with those served by the sense amplifiers. In the \textit{write mode}, the verify logic, together with the write pulse shaper logic, can serve P\&V-based write operations. Table \ref{tab:rr} summarizes our findings.

\begin{table}[h!]
	\renewcommand{\arraystretch}{0.8}
	\centering
	{\fontsize{6}{10}\selectfont
		\begin{tabular}{|c|c|c|c|c|}
			\hline
			\textbf{Operating} & \multicolumn{2}{|c|}{\textbf{Write driver circuit}} & \multirow{2}{*}{\textbf{Sense Amplifier}} & \textbf{Operations}\\\cline{2-3}
			\textbf{Modes} & \textbf{Pulse shaper logic} & \textbf{Verify logic} & & \textbf{Enabled}\\
			\hline
			\hline
			write & $\surd$ & $\surd$ & $\times$ & one write request \\
			decoupled &  $\times$ & $\surd$ & $\surd$ & two read requests\\
			\hline
	\end{tabular}}
	\caption{Two operating modes of the write driver circuit introduced using our circuit modifications.}
	\label{tab:rr}
\end{table}
This is \textit{precisely} the read-read parallelism we seek in a PCM bank to resolve a \textbf{read-read} bank conflict, where the two conflicting requests are to two different partitions within the bank. 
To enable this parallelism,
we introduce two \emph{new} PCM commands:
\begin{itemize}
	\item \texttt{\underline{DECOUPLE} (D):} set \texttt{M4 = OFF}.
	\item \texttt{\underline{READ-WITH-READ} (RWR):} connect the sense amplifiers and verify logic of the write drivers to the two decoded partitions.
\end{itemize}

To facilitate arbitration of the data bus for transferring two sets of read data back to the memory controller, we introduce two more transistors \texttt{M5} and \texttt{M6}. In the write mode, \texttt{M5 = OFF} and \texttt{M6 = ON}. In the decoupled mode, the memory controller first retrieves the sense amplifiers' data in 8 cycles. It then sets \texttt{M5 = ON} and \texttt{M6 = OFF} in the next cycle, which enables the verify logic's data to be retrieved in the following 8 cycles. The data transfer latency from PCM to the memory controller is therefore \ineq{8+1+8= 17} cycles.
For this purpose, we introduce our final new PCM command:
\begin{itemize}
	\item \texttt{\underline{TRANSFER} (T):} sets \texttt{M5 = ON} and \texttt{M6 = OFF}.
\end{itemize}

\begin{figure}[h!]
	\centering
	\centerline{\includegraphics[width=0.79\columnwidth]{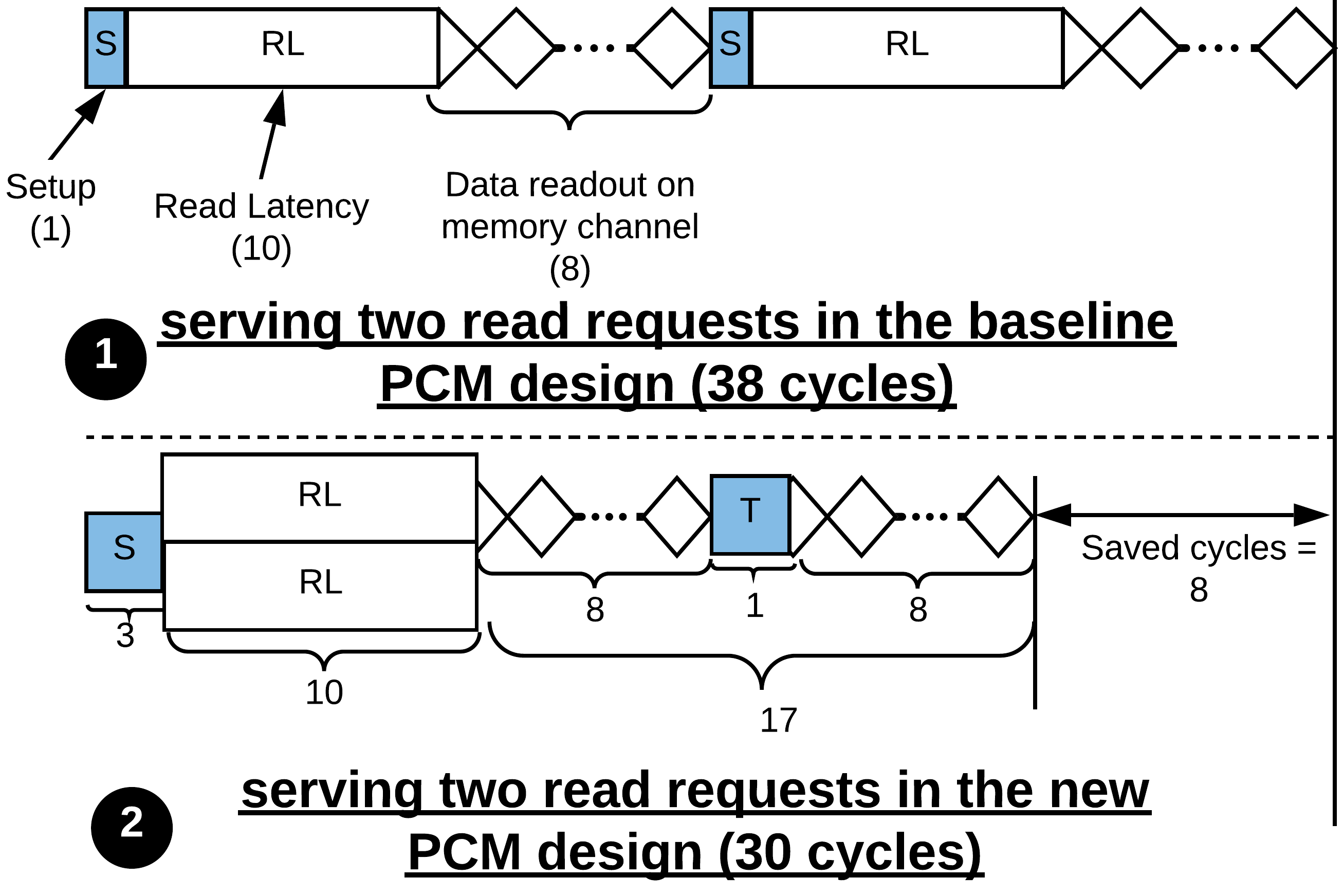}}
	\caption{Performance in the baseline (\ding{182}) and our PCM design (\ding{183}) when scheduling two read requests to different partitions in the same bank.}
	\label{fig:savings_rr}
\end{figure}

Figure \ref{fig:savings_rr} compares how two read requests are scheduled in the baseline PCM design (\ding{182}) with our PCM design (\ding{183}), where the memory controller exploits the PCM bank's partition-level parallelism. Following are the respective command sequences. 

\begin{varwidth}[t]{0.5\textwidth}
	\footnotesize{Baseline PCM:}
	\begin{itemize}
		\item \texttt{ACTIVATE address in \ineq{i}}
		\item \texttt{READ}
		\item \texttt{PRECHARGE}
		\item \texttt{ACTIVATE address in \ineq{j}}
		\item \texttt{READ}
		\item \texttt{PRECHARGE}
		\item[--] Service time = 19 + 19
		\item[] = 38 cycles
	\end{itemize}
\end{varwidth}
\hspace{2em}
\begin{varwidth}[t]{.5\textwidth}
	\footnotesize{Proposed PCM:}
	\begin{itemize}
		\item \texttt{ACTIVATE address in \ineq{i}}
		\item \texttt{ACTIVATE address in \ineq{j}}
		\item \texttt{DECOUPLE}
		\item \texttt{READ-WITH-READ}
		\item \texttt{TRANSFER}
		\item \texttt{PRECHARGE}
		\item[--] Service time = 1 + 1 + 1 + 10 + 
		\item[ ] 8 + 1 + 8 = 30 cycles
		\item[ ]
	\end{itemize}
\end{varwidth}

In the baseline PCM design, each \ineq{\texttt{A}-\texttt{R}-\texttt{P}} takes 19 cycles. The total service latency is therefore, 38 cycles to serve two read requests serially from partitions \ineq{i} and \ineq{j}. 
In our PCM design, the memory controller issues the following commands: \ineq{\texttt{A}-\texttt{A}-\texttt{D}-\texttt{RWR}-\texttt{T}-\texttt{P}}. In our new design, \texttt{RWR} takes 10 cycles, during which the read latency (RL) of the two read requests are overlapped. The total service latency is therefore \ineq{1+1+1+10+17 = 30} cycles, considering 8+1+8 = 17 cycles for the data transfer from PCM to the memory controller as shown in Figure \ref{fig:savings_rr} (\ding{183}).






%% file: sections/exploit_palp.tex
This section describes our new memory access scheduling policy to exploit each PCM bank's partition-level parallelism, which we describe how to enable in Section \ref{sec:enable_PALP}.
\subsection{High-level overview}
The key idea of our scheduling policy is to maximize partition-level parallelism as long as the running average power limit (RAPL) is not violated and a request does not become delayed too much.

At a high level, the memory controller maintains a read-write queue (rwQ) to store 
PCM requests. We implement rwQ as a FIFO. After scheduling a request from the rwQ, the memory controller checks to see if there is any outstanding request that can be scheduled exploiting the PCM bank's partition-level parallelism. If so, the request is scheduled \textit{along with} the ongoing request, and the memory timing parameters are set accordingly. Otherwise, the memory controller selects the oldest request in the rwQ to be scheduled \textit{after} completing the ongoing one.

\begin{figure}[h!]
	\centering
	\centerline{\includegraphics[width=0.79\columnwidth]{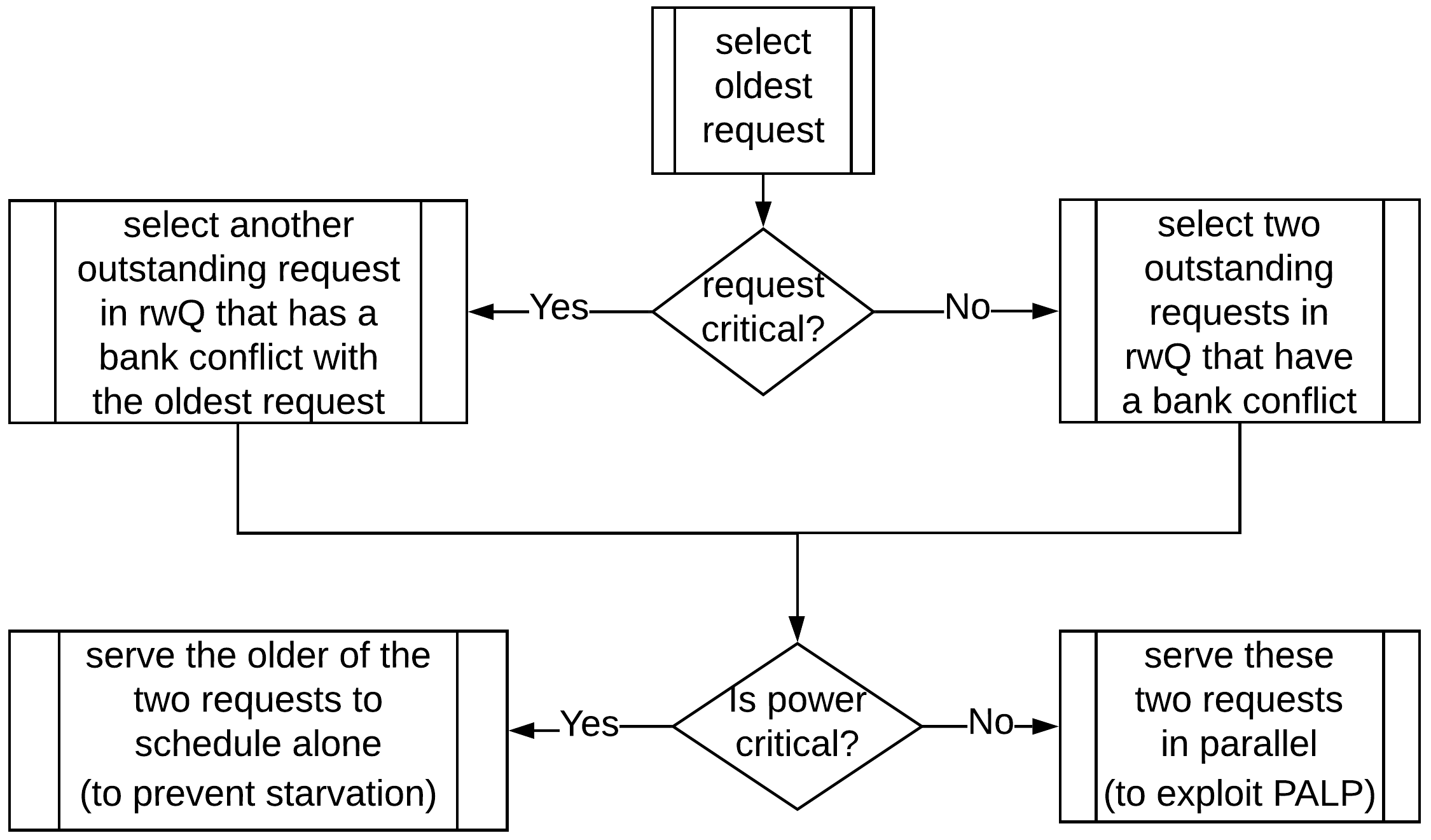}}
	\caption{Flowchart describing our new memory access scheduling policy.}
	\label{fig:policy_transitions}
\end{figure}

Figure \ref{fig:policy_transitions} summarizes the flowchart of our new memory access scheduling policy.

\subsection{Detailed design} 
To explain our new memory access scheduling policy, we introduce the following notation
\begin{align*}
    r_l^p (w_l^p) &\text{: PCM read (write) access to wordline } l \text{ in partition } p\\
    th_b & \text{: starvation threshold}\\
    o(x) & \text{: number of memory clock cycles for which}\\
    & \text{ request } x \text{ is outstanding in } rwQ \\
    P_\text{SA} & \text{: power consumption of all sense amplifiers in a bank}\\
    P_\text{WD} & \text{: power consumption of all write drivers in a bank}\\
    N & \text{: total number of memory clock cycles elapsed}\\
    P & \text{: running average power consumption}\\
    RAPL & \text{: running average power limit \cite{david2010rapl} for the PCM device}
\end{align*}
We estimate the power consumed to resolve a read-read (R-R) and read-write (R-W) bank conflict as follows:
\begin{equation}
\label{eq:power_estimate}
\scriptsize P_\text{est}^\text{R-R} = \frac{N*P + 30*P_{SA} + 30*P_{WD}}{N + 30},  P_\text{est}^\text{R-W} = \frac{N*P + 48*P_{SA} + 48*P_{WD}}{N + 48}
\end{equation}
The timing parameters are as follows: 30 cycles for resolving the read-read (R-R) conflict, and 48 cycles for the read-write (R-W) conflict (see also Table \ref{tab:simulation_parameters}).

The pseudo-code of our new memory access scheduling policy is shown in Algorithm \ref{alg:palp}.

\begin{algorithm}[t]
	\scriptsize{
		\texttt{next-request} $= \displaystyle \max_{x \in rwQ} o(x)$ \tcc*[r]{select the oldest request in the rwQ}
		\If(\tcc*[f]{if the request \texttt{next-request} is not critical, i.e., the number of clock cycles for which it is outstanding in the $rwQ$ is within the backlogging threshold $th_b$}){$o(\texttt{next-request}) < th_b$}
		{
		\texttt{next-request} = select the oldest request in the $rwQ$ that has bank conflict
		}
		Let \texttt{next-request} = $a_l^p$\;
		$W^p$ = set of $n_w$ write requests in $rwQ$ to partition $p$\; 
		$R^p$ = set of $n_r$ read requests in $rwQ$ to partition $p$\; 
		\tcc*[r]{Find a request that can be concurrently scheduled with $a_l^p$}
		\uIf{$a_w^p$ is a write request}
		{
		    \texttt{concurrently-scheduled-request} $= \displaystyle \max_{i=1,\dots n_r} o(r_i^p)~\mid r_i^p \in R^p$\ \tcc*[r]{select the oldest request in the set $R^p$}
		    $P = P_\text{est}^{\text{R-W}}$ \tcc*[r]{calculate power for concurrent request scheduling $P_\text{est}^{\text{R-W}}$ using Equation \ref{eq:power_estimate}}
		}
		\Else(\tcc*[f]{$a_w^p$ is a read request}){
		    \tcc*[r]{prioritize scheduling the write request because resolving read-write bank conflicts achieves more performance improvement compared to resolving read-read bank conflicts (determined emperically).}
		    \texttt{concurrently-scheduled-request} $= \displaystyle \max_{i=1,\dots n_w} o(w_i^p)~\mid w_i^p \in W^p$\ \tcc*[r]{select the oldest request in the set $W^p$}
		    $P = P_\text{est}^{\text{R-W}}$ \tcc*[r]{calculate power for concurrent request scheduling $P_\text{est}^{\text{R-W}}$ using Equation \ref{eq:power_estimate}}
		    \If(\tcc*[f]{there are no write requests to partition $p$ that can be concurrently scheduled with $a_l^p$}){$W^p = \emptyset$}
		    {
		        \tcc*[r]{select a read request that can be concurrently scheduled and calculate the power overhead}
		        \texttt{concurrently-scheduled-request} $= \displaystyle \max_{i=1,\dots n_r} o(r_i^p)~\mid r_i^p \in R^p$\ \tcc*[r]{select the oldest request in the set $R^p$}
		        $P = P_\text{est}^{\text{R-R}}$ \tcc*[r]{calculate power for concurrent request scheduling $P_\text{est}^{\text{R-R}}$ using Equation \ref{eq:power_estimate}}
		    }
		}
		\uIf(\tcc*[f]{estimated power consumption of concurrent schedule is within the RAPL limit}){$P\leq RAPL$}
		{
		    \texttt{next-request-to-schedule} = \{$a_l^p$,\texttt{concurrently-scheduled-request}\}
		}
		\Else{
		    \texttt{next-request-to-schedule} = \{$a_l^p$\}\tcc*[r]{select the oldest request to schedule alone}
		}
	}
	\caption{Our new memory access scheduling policy.}
	\label{alg:palp}
\end{algorithm}

\subsection{Significance of \tech{}'s scheduling policy}
\label{sec:fcfs_limitations}
We provide some intuition, via an example, as to why (1) enabling each PCM bank's partition-level parallelism is not sufficient to significantly improve performance, unless there is a scheduling policy that explicitly exploits such parallelism, and (2) why \tech{}'s scheduling policy outperforms the standard FCFS policy \cite{RixnerISCA2000}, which is commonly used by many PCM memory controllers.

\begin{figure}[h!]
	\centering
	\centerline{\includegraphics[width=0.99\columnwidth]{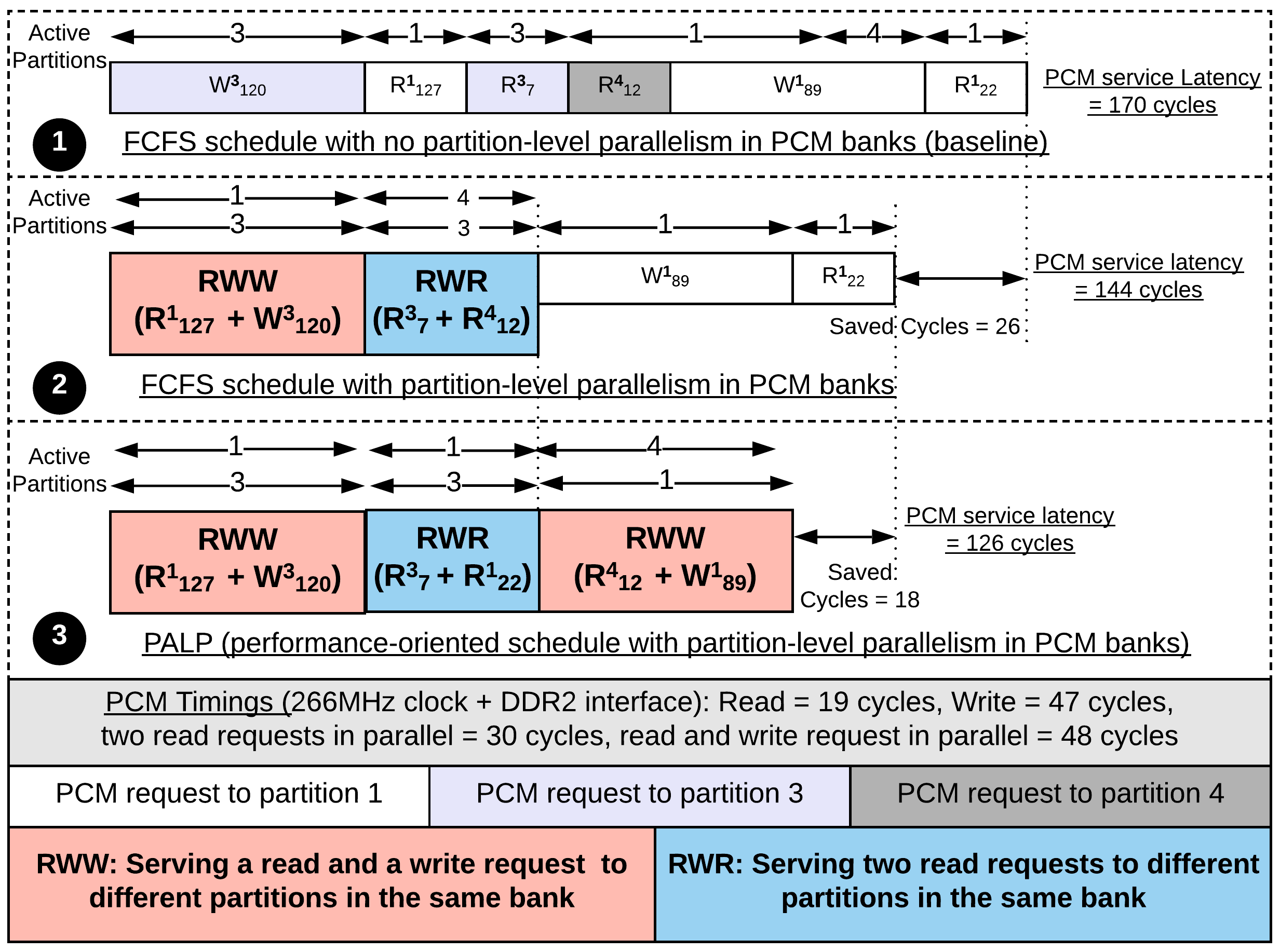}}
	\caption{Example request schedules using \ding{182} the FCFS schedule of the Baseline \cite{ArjomandISCA16}, \ding{183} FCFS schedule exploiting PCM bank's partition-level parallelism, and \ding{184} \tech{}'s new memory access scheduling policy.}
	\label{fig:schedule_motivation}
\end{figure}

Figure \ref{fig:schedule_motivation} illustrates an example showing how the memory controller schedules six PCM requests to the same bank.
In (\ding{182}) we illustrate the FCFS policy of the Baseline \cite{ArjomandISCA16}, where no more than \emph{one} partition is active at any time.
Using the timing parameters listed at the bottom of this figure, the total PCM service latency is 170 cycles.

In (\ding{183}) we illustrate the FCFS policy with partition-level parallelism in PCM banks. 
The PCM request \ineq{R^1_{127}} is scheduled with \ineq{W^3_{120}} by issuing the new PCM command \texttt{RWW} using partitions 1 and 3. This resolves the read-write bank conflict. The PCM request \ineq{R^4_{12}} is scheduled with \ineq{R^3_7} by issuing the new PCM command \texttt{RWR} using partitions 4 and 3. This resolves the read-read bank conflict. 
We note that requests \ineq{W^1_{89}} and \ineq{R^1_{22}} are both to partition 1, and therefore scheduled serially.
In this example, more than one partition can be active, which enables the total PCM service latency to go down to 144 cycles, a reduction of 15.3\% over the baseline.

Finally, in (\ding{184}) we illustrate one example schedule obtained using \tech{}'s new memory access scheduling policy. The memory controller re-orders PCM requests in the read-write queue to maximize partition-level parallelism. We observe 1) request  \ineq{R^1_{127}} is scheduled with \ineq{W^3_{120}}, and request \ineq{R^4_{12}} is scheduled with \ineq{W^1_{89}} by issuing two \texttt{RWW} PCM commands and 2) request \ineq{R^3_{7}} is scheduled with \ineq{R^1_{22}} by issuing a \texttt{RWR} PCM command.
This reduces the total PCM service latency to 126 cycles, a further savings of 12.5\% compared to \ding{184}.

Overall, \tech{} improves performance by 25.8\% over the baseline \ding{183} in this example. In Section \ref{sec:results}, we report \tech{}'s application-level performance improvement for all our evaluated workloads.

%% file: sections/evaluation_methodology.tex
To evaluate \tech{}, we develop a full-system simulator with the following components.
\begin{itemize}
	\item \hl{Gem5 \cite{binkert2011gem5} simulator frontend to simulate an ARMv8-A (aarch64) system \cite{armstrong2019isa} with 8 cores.}
	\item We use a hybrid DRAM-PCM memory system. DRAMPower \cite{chandrasekar2012drampower} is used to estimate its power consumption. 
	\item  \hl{In-house cycle-level PCM simulator for 8GB, 16GB, and 32GB PCM with DDR4 interface.} This is based on Ramulator \cite{kim2016ramulator}, a cycle-accurate main memory simulator. Power and latency parameters are based on IBM's 20nm PCM \cite{lung2016double}, \hl{with DDR4 interface parameters \cite{jedec2012jedec}}.
	Our simulator is available for download at \cite{simulator}.
\end{itemize}

Table \ref{tab:simulation_parameters} shows our simulation parameters.

\begin{table}[h!]
	\centering
	{\fontsize{8}{10}\selectfont
		\begin{tabular}{lp{6.5cm}}
			\hline
			Processor & \hl{8 cores, 3 GHz, out-of-order}\\
			\hline
			L1-I/D cache & \hl{Private 64KB per core, 4-way}\\
			\hline
			L2 cache & \hl{Private 512KB per core, 8-way}\\
			\hline
			L3 cache & \hl{Shared 4MB, 16-way}\\
			\hline
			eDRAM cache & \hl{Shared, writes-only, and on-chip}\\
								  & \hl{4MB (default), 8MB, 16MB, 32MB, and 64MB eDRAMs are evaluated}\\
			\hline
			Main memory & \hl{8GB (default), 16GB, and 32GB PCMs are evaluated} \\
								 & (4 channels, 4 ranks/channel, 8 banks/rank.)\\
								 & \hl{Memory interface = DDR4}\\
								 & \hl{Memory clock = 256MHz}\\
			& \texttt{A-R-P} / \texttt{A-W-P} = 19 cycles / 47 cycles\\
			& tWR / RL = 35 cycles / 10 cycles\\
			& \texttt{A-RWW-P} / \texttt{A-RWR-P} = 48 cycles / 30 cycles\\
			\hline
	\end{tabular}}
	\caption{Major simulation parameters.}
	\label{tab:simulation_parameters}
\end{table}


\subsection{Evaluated techniques}
We evaluate the following techniques:
\begin{itemize}
	\item \emph{Baseline:} The Baseline technique \cite{ArjomandISCA16} maximizes PCM performance by serving multiple requests in parallel across all banks. It uses the FCFS policy and does \textit{not} exploit partition-level parallelism.
	\item \emph{MultiPartition:} \mc{The MultiPartition technique \cite{zhou2016efficient} can resolve the read-write bank conflicts in PCM by exploiting the presence of partitions in a bank.} The original design of \cite{zhou2016efficient} uses the FCFS policy, which provides very small benefit over the Baseline (according to our evaluations). For a fair comparison with \tech{}, we implemented out-of-order scheduling for this technique. This scheduling policy explicitly prioritizes requests that exploit read-write parallelism in PCM. Although \cite{zhou2016efficient} does not consider memory interface timings, we implement the MultiPartition technique with the DDR4 interface to fairly and accurately estimate its performance.
	\item \emph{\tech{}:} Our mechanism enables each PCM bank's partition-level parallelism, and uses our new memory access scheduling policy to optimize performance by prioritizing requests that can exploit such partition-level parallelism. \mc{\tech{} can resolve both read-write and read-read bank conflicts.}
\end{itemize}
The address mapping for all our evaluated systems is based on Micron's DDR4 Datasheet \cite{micron2014sdram}. We show an example of how a memory address is mapped in PCM with 4 channels, 4 ranks/channel, 8 banks/rank, 8 partitions/bank: [36:35]=rank, [34:23]=row, [22:14]=column, [13: 11]=partition, [10:8]=bank, [7:6]=channel, [5:0]=byte within a cache line.
We also evaluated various different forms of address interleaving, ranging from cache block interleaving to row interleaving \cite{kim2010atlas}. Even though exact numerical benefits differ, our mechanism works and improves performance for all evaluated address mappings.

\subsection{Evaluated workloads}
Table \ref{tab:spec} reports the evaluated workloads. 
These workloads were chosen because they have at least 1 memory access per 1000 instructions out of the 64MB on-chip eDRAM cache. \hl{Although SPEC CPU2017 workloads are commonly used for evaluating high performance systems, recent works suggest that these workloads are also representative of many emerging applications that are regularly enabled by users on their mobile phones \cite{panda2018wait}.}

\begin{table}[h!]
	\centering
	{\fontsize{8}{10}\selectfont
		\begin{tabular}{p{5cm}p{3cm}}
			\hline
			\textbf{\hl{MiBench workloads \cite{guthaus2001mibench}}} & \textbf{}\\
			\multicolumn{2}{p{8cm}}{\hl{tiff2rgba, jpeg\_decode, tiffdither, susan\_smoothing, typeset}}\\
			\hline
			\textbf{SPEC CPU2017 workloads \cite{bucek2018spec}} & \textbf{}\\
			\multicolumn{2}{p{8cm}}{cactusBSSN, bwaves, roms, parset, xz}\\
			\hline
			\textbf{Mixed (parallel) applications} & \textbf{}\\
			\multicolumn{2}{p{8cm}}{\hl{\textbf{AI-1} (4 copies each of deepsjeng, leela), \textbf{AI-2} (4 copies each of mcf, exchange2), \textbf{Visualization-1} (4 copies each of povray, blender), \textbf{Visualization-2} (4 copies each of povray, imagick), \textbf{Scientific} (4 copies each of cactusBSSN, bwaves)}}\\
			\hline
	\end{tabular}}
	\caption{Evaluated workloads.}
	\label{tab:spec}
\end{table}

\subsection{Figures of Merit}
We report the following figures of merit in this work to evaluate different mechanisms:

\begin{itemize}
    \item[1.] \emph{Execution Time:} The time it takes to finish a workload.
    \item[2.] \emph{Queuing Delay:} The total number of memory cycles spent by a request in the rwQ, averaged over all PCM requests. The delay of each request is measured as the time difference between when a request is inserted in the queue and the time when it is scheduled to PCM.
	\item[3.] \emph{Access Latency:} The sum of queuing delay and the PCM service latency, averaged over all PCM requests. 
	\item[4.] \emph{PCM Power Consumption:} The total power consumed for activating partitions and peripheral structures within PCM banks.
\end{itemize}

%% file: sections/results.tex
\subsection{Overall system performance}
\label{sec:performance_improvement}
\begin{figure}[h!]
	\centering
	\centerline{\includegraphics[width=0.99\columnwidth]{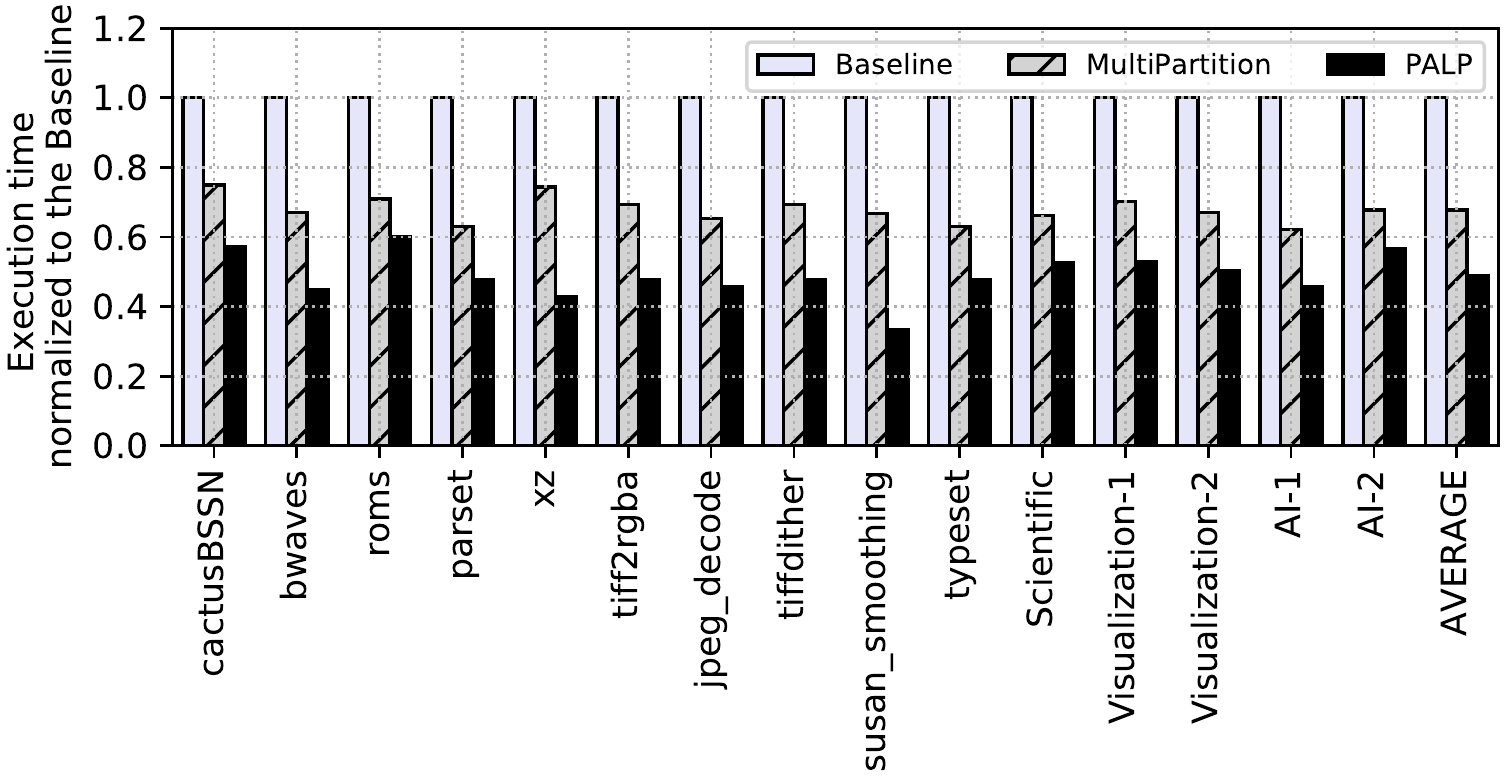}}
	\vspace{-10pt}
	\caption{\hl{Execution time with \tech{}, normalized to Baseline for our default configuration of 4MB eDRAM cache and 8GB PCM.}}
	\label{fig:sppedup}
\end{figure}

\hl{Figure \ref{fig:sppedup} reports the execution time of each of our workloads for each of our evaluated systems normalized to the Baseline system. The simulator is configured for the default settings of 4MB eDRAM cache and a 8GB PCM. We make the following two observations.}

\hl{\textbf{First}, \mpat{} has higher performance than the Baseline (32\% lower average execution time). This improvement is because \mpat{} resolves read-write bank conflicts in PCM. 
\textbf{Second}, \tech{} has the highest performance among all the three evaluated systems (\tech{} has 51\% lower average execution time than the Baseline, and \perf{} lower average execution time than \mpat{}). This performance improvement is because 1) \tech{} resolves both read-read and read-write bank conflicts, and 2) \tech{}'s memory access scheduling policy is optimized to maximize both read-read and read-write partition-level parallelism to achieve higher performance.
}
\subsection{Queuing delay}
\begin{figure}[h!]
	\centering
	\centerline{\includegraphics[width=0.99\columnwidth]{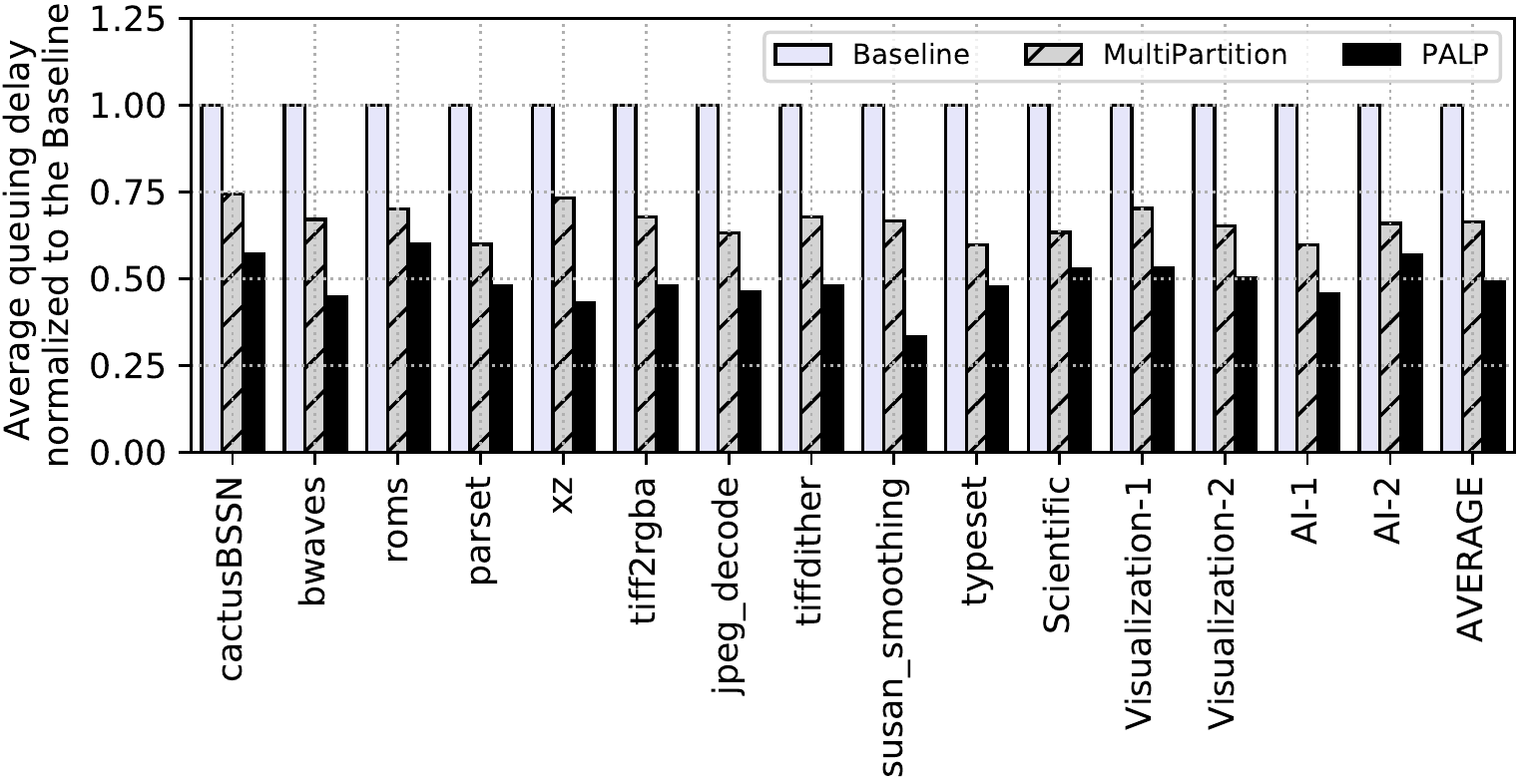}}
	\vspace{-10pt}
	\caption{\hl{Queuing delay of \tech{}, normalized to Baseline.}}
	\label{fig:wait}
\end{figure}

\hl{Figure \ref{fig:wait} reports the queuing delay of each of our workloads for each of our evaluated systems normalized to the Baseline system. The simulator is configured for the default settings of 4MB eDRAM cache and a 8GB PCM. We make the following two observations.}

\hl{
	\textbf{First,} the average queuing delay of \mpat{} is 34\% lower than Baseline. This is because \mpat{} reduces the average delay of the outstanding write requests by scheduling them concurrently with read requests that are to different partitions within PCM banks (exploiting read-write parallelism). 
	\textbf{Second}, the average queuing delay of \tech{} is the lowest (52\% lower than the Baseline and 26\% lower than \mpat{}). This reduction is because \tech{} also reduces the queuing delay of read requests by scheduling them concurrently with those that are to different partitions within PCM banks (exploiting read-read parallelism).
}

\subsection{Access latency}
\begin{figure}[h!]
	\centering
	\centerline{\includegraphics[width=0.99\columnwidth]{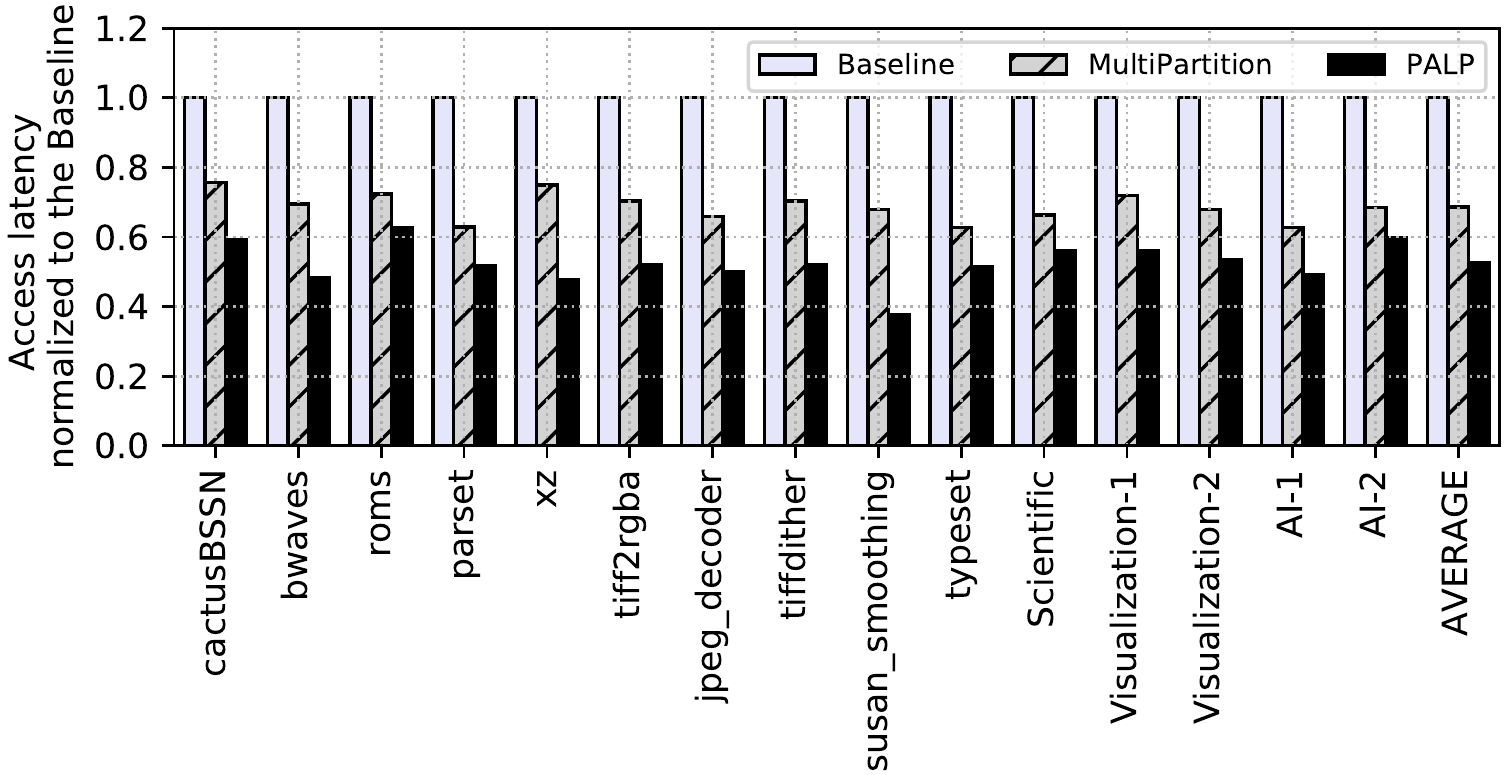}}
	\vspace{-10pt}
	\caption{\hl{Access latency with \tech{}, normalized to Baseline.}}
	\label{fig:access}
\end{figure}

\hl{Figure \ref{fig:access} reports the access latency of each of our workloads for each of our evaluated techniques normalized to the Baseline, using the default settings of our simulator.
We make the following two observations.}

\hl{
	\textbf{First}, the average access latency of \mpat{} is 31\% lower than the Baseline. This reduction is a result of performance improvement due to exploiting read-write partition-level parallelism in PCM, as we have discussed in Section \ref{sec:rw}. 
	\textbf{Second}, the average access latency of \tech{} is the lowest among all the three systems (47\% lower than the Baseline, and \alat{} lower than \mpat{}). This reduction is due to the significant reduction of the PCM service latency achieved by exploiting both read-read and read-write partition-level parallelism in PCM banks. 
}

\subsection{PCM power consumption}
\begin{figure}[h!]
	\centering
	\centerline{\includegraphics[width=0.99\columnwidth]{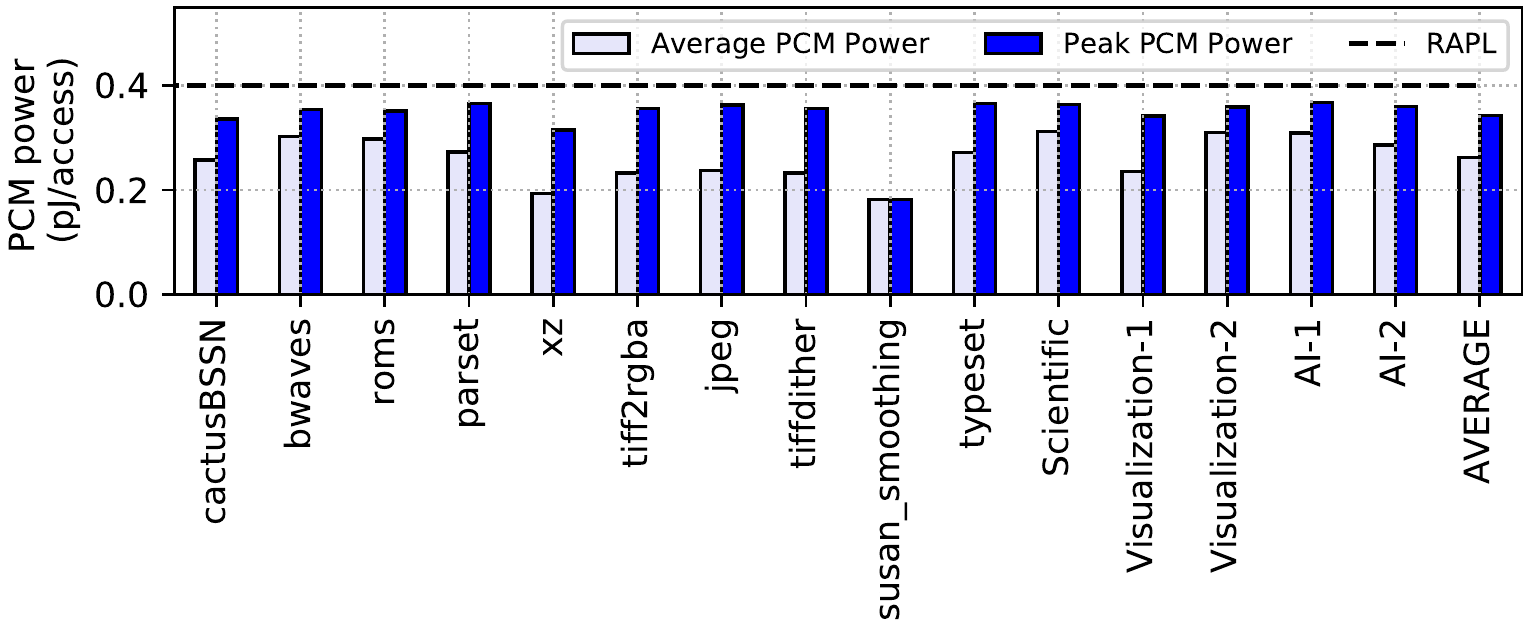}}
	\caption{Average and peak power consumption of PCM with \tech{}.}
	\label{fig:power}
\end{figure}

\hl{Figure \ref{fig:power} reports the average and peak power consumption of \tech{} for each of our workloads. The simulator is configured for the default settings of 4MB eDRAM cache and a 8GB PCM. We also report the RAPL limit, which is 0.4pJ per access. This limit is what is specified in the PCM datasheet \cite{lung2016double}. We make the following two observations.}

	\textbf{First}, both average and peak PCM power consumption of \tech{} is within the RAPL limit. The average PCM power is at least 0.08pJ/access lower than the RAPL limit, while the peak PCM power is at least 0.03pJ/access lower. Our memory access scheduler explicitly estimates the increase in power consumption when more than one partition is active to exploit partition-level parallelism, and serializes requests to PCM anytime the RAPL limit is estimated to be exceeded. 
	\textit{None} of the prior works take 
	power consumption into account while exploiting PCM bank's partition-level parallelism. As a result, the RAPL limit cannot be guaranteed in these works.
	\textbf{Second}, our technique allows the exploration of performance and power trade-offs. We observe that reducing the RAPL limit to 0.35pJ/access will not hurt performance when our technique is employed.
	In fact, system designers can potentially use our technique to estimate the RAPL needed to achieve a desired performance target and distribute the surplus power budget to other system components.

\subsection{Latency, power, and area overheads}
We evaluate the latency and power overhead of \tech{} against the Baseline PCM design using SPICE simulations \cite{emam1997spice} with 20nm technology files from \cite{sinha2012exploring}. For the Baseline design, we use IBM's PCM architecture \cite{lung2016double}. In evaluating our design overheads, we use the industry standard low standby power (LSTP) multi-gate transistors. We also use copper interconnect parasitics \cite{schuddinck2012standard} in our SPICE simulation to model wire delays. Finally, we modeled process variation using the guidelines presented in \cite{xiong2003sensitivity}. In Table \ref{tab:compare_design_overheads}, we report the latency and power overheads of \tech{}.

\begin{table}[h!]
	\renewcommand{\arraystretch}{0.8}
	\centering
	{\fontsize{6}{10}\selectfont
		\begin{tabular}{|c|c|c|c|c|}
			\hline
			\multirow{2}{*}{\textbf{Design}} & \textbf{Technology} & \textbf{Vdd}  & \textbf{Critical Path Delay}  & \textbf{Power}\\
															 & \textbf{Type}		    & \textbf{(V)}   & \textbf{(ps)}     & \textbf{(pJ/access)}\\
			\hline
			\hline
			Baseline \cite{lung2016double} & Double-gate & 0.9 & 1159.2 & 0.311\\
			Proposed \tech{} 					  & Double-gate & 0.9 & 1453.2 & 0.364\\
			\hline
	\end{tabular}}
	\caption{\hl{Design parameters for the Baseline \cite{lung2016double} vs. \tech{} for a single peripheral structure consisting of a sense amplifier and a write driver.}}
	\label{tab:compare_design_overheads}
\end{table}

\hl{We observe that due to the new circuits that we introduce, the critical path delay has increased by 25.3\%. However, the clock cycle time is 3.9ns for 256MHz rated memory clock, which is much higher than the critical path delay of 1453.2ps. We believe that the introduced logic is unlikely to create setup or hold violations at this rated clock frequency.}

\hl{Our design also increases the power consumption by 17\% over the Baseline design of a single peripheral structure consisting of a sense amplifier and our new write driver. 
}

\hl{Finally, we observe that \tech{} has an area overhead of 1.15\% compared to the area of each peripheral structure. This overhead is negligible compared to the area of a 1GB PCM bank, which is \ineq{9.43\times6.30} mm\ineq{{}^2} at \ineq{20}nm technology node.}

\subsection{Effect of \tech{} with different PCM capacities}
\begin{figure}[h!]
	\centering
	\centerline{\includegraphics[width=0.99\columnwidth]{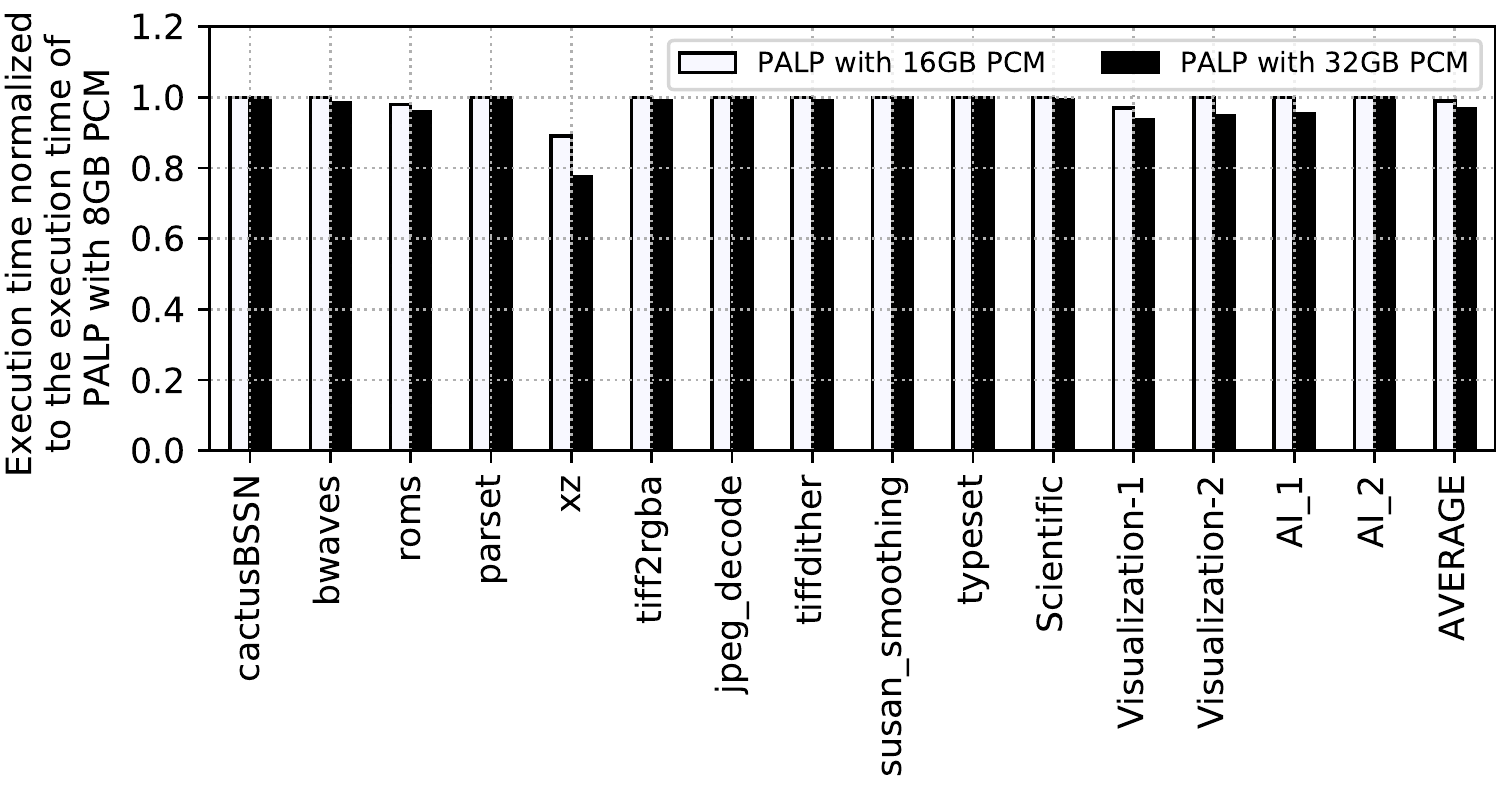}}
	\vspace{-10pt}
	\caption{Execution time with \tech{}, for different PCM capacities, normalized to the default configuration of \tech{} with 8GB PCM.}
	\label{fig:pcm_size}
\end{figure}

Figure \ref{fig:pcm_size} reports the execution time of each of our workloads with \tech{} using PCM capacities of 16GB and 32GB, normalized to the default configuration using \tech{} with 8GB PCM.
The eDRAM capacity is configured to 4MB for all these configurations. We make the following two observations from our study.

	\textbf{First}, for most workloads, we observe a very little performance improvement when the PCM capacity is increased from 8GB to 32GB. This is because these workloads have small working sets, for which a 8GB PCM is sufficient. 
	\textbf{Second}, for xz we observe a significant performance improvement when we increase the PCM capacity to 16GB and 32GB due to xz's large working set. For this workload, the higher the number of PCM banks, the better the performance of \tech{}. This is because \tech{} can exploit more parallelism in more partitions in PCM that exist in more banks. 

\subsection{Effect of different eDRAM capacities}
\begin{figure}[h!]
	\centering
	\centerline{\includegraphics[width=0.99\columnwidth]{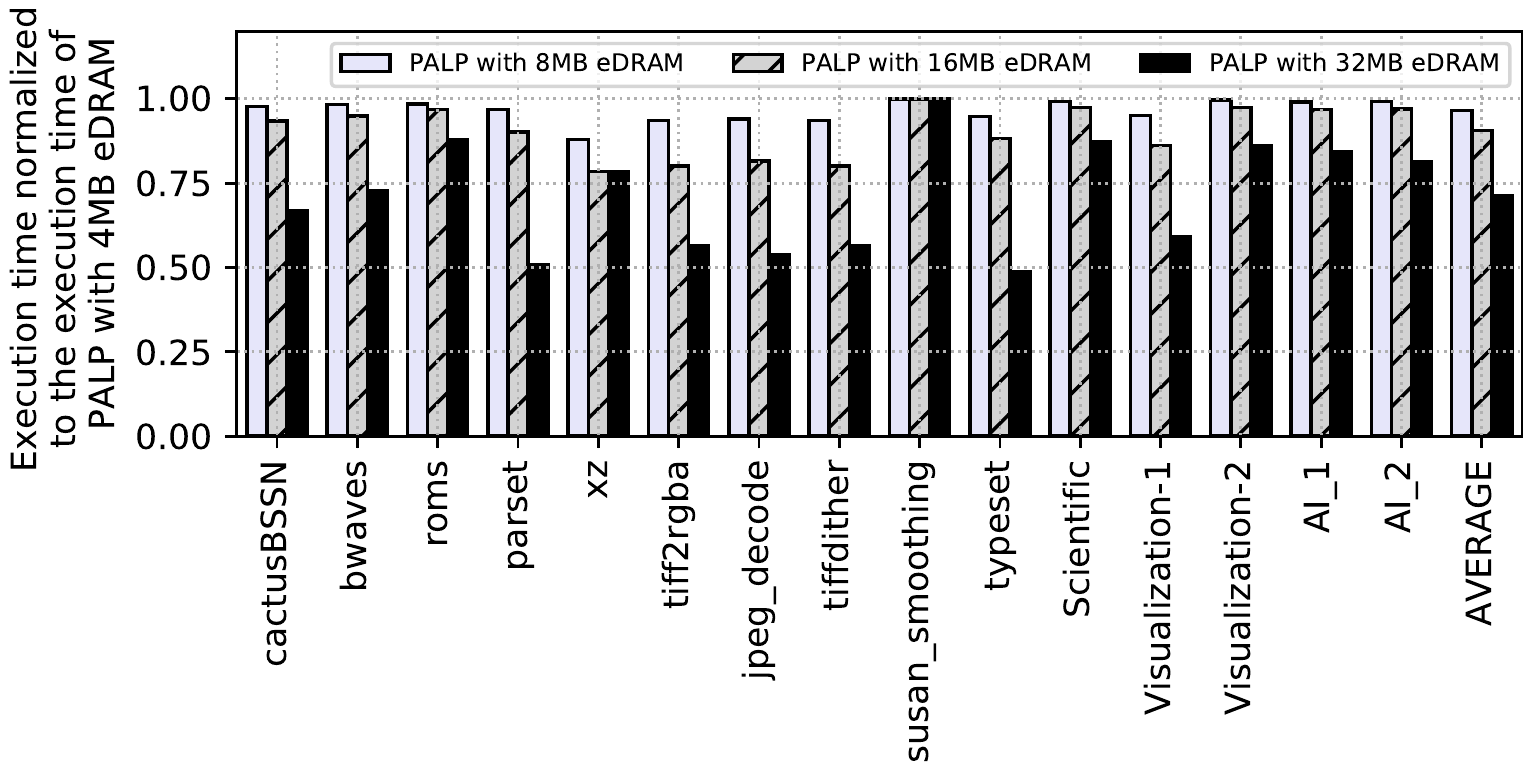}}
	\caption{Execution time with \tech{}, for different eDRAM capacities, normalized to the default configuration of \tech{} with 4MB eDRAM.}
	\label{fig:edram_size}
\end{figure}

Figure \ref{fig:edram_size} reports the execution time of each of our workloads with \tech{}, normalized to the default configuration using \tech{} with 4MB eDRAM cache. We report results for \tech{} with eDRAM capacity of 8MB, 16MB, and 32MB. The PCM capacity is configured to 8GB. We make the following two observations.

	\textbf{First}, for most workloads, as we increase the capacity of the eDRAM cache we observe a significant improvement in performance (lower execution time). This is because with a larger eDRAM cache capacity, more write requests are absorbed in the eDRAM, leaving only the read requests to be queued in the rwQ. This impacts the execution time in the following two ways: 1) the latency to service long write requests is reduced, and 2) the memory controller now has more flexibility to exploit read-read parallelism from the outstanding read requests. 
	\textbf{Second}, for susan\_smoothing, we observe a marginal change in performance when we increase the eDRAM capacity to 32MB. This is because there are only a small number of write requests in this workload to begin with, and therefore the workload's performance is insensitive to the size of the eDRAM cache, which buffers only write requests.

\subsection{\hl{Effect of different interface timings}}
\begin{figure}[h!]
	\centering
	\centerline{\includegraphics[width=0.99\columnwidth]{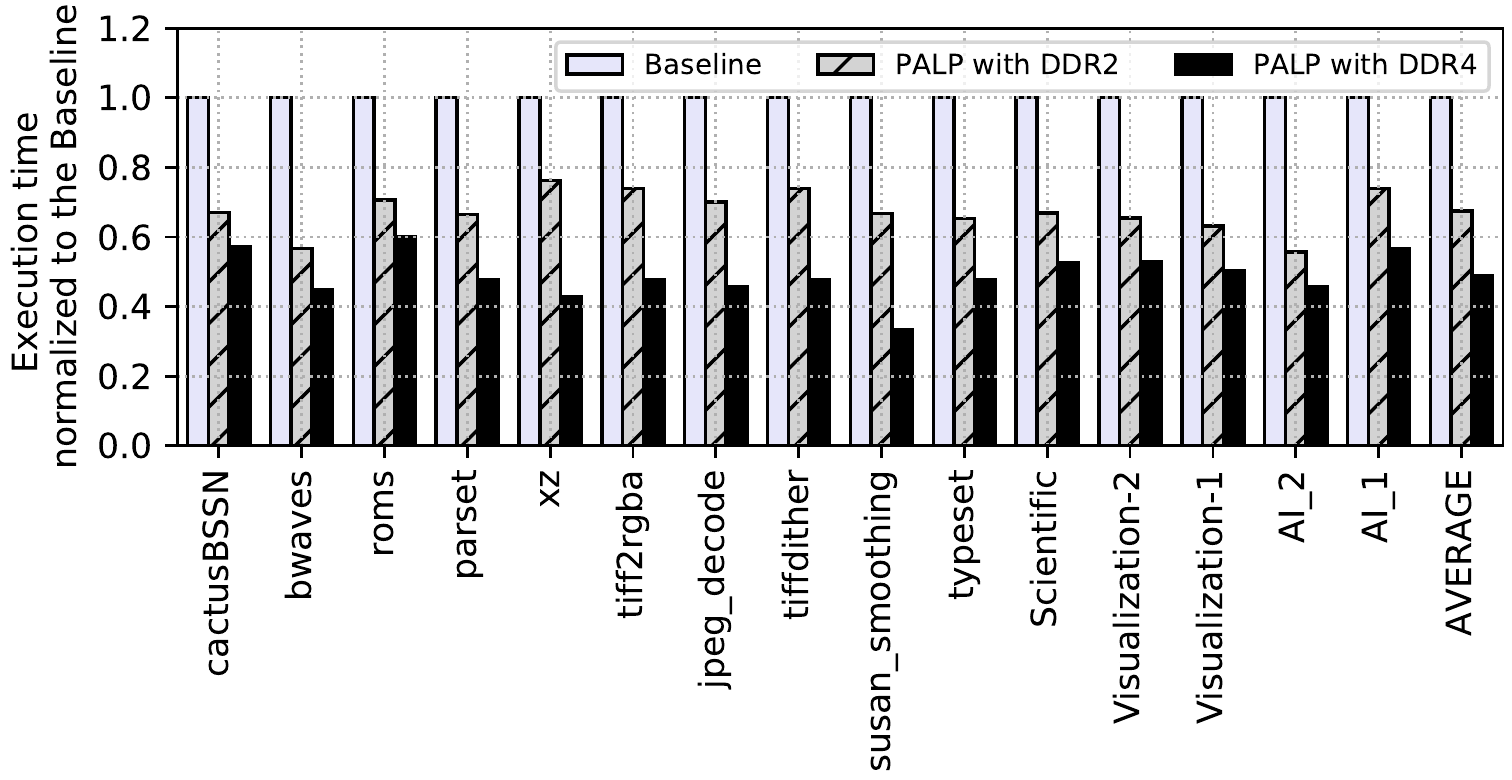}}
	\caption{Execution time with \tech{}, for different DDRx interfaces, normalized to the Baseline.}
	\label{fig:interface}
\end{figure}

\hl{Figure \ref{fig:interface} reports the execution time of each of our workloads for \tech{} with DDR2 and DDR4 interfaces, normalized to the Baseline using the default settings of our simulator. 
We make the following two observations.}

    \textbf{First}, the performance of \tech{} with DDR2 and DDR4 interfaces are both better than the Baseline. The average execution time of \tech{} with the DDR2 interface is 33\% lower than the Baseline, while that with the DDR4 interface is 51\% lower than the Baseline.
	\textbf{Second}, we observe that the execution time of \tech{} decreases when the memory interface is changed from DDR2 to DDR4. The average performance improves by 27\% when DDR4 interface is used. This increase is because the data transfer rate is doubled in DDR4 compared to DDR2. A complete exploration of all DRAM standards is a vast undertaking, and is beyond the scope of this work (see, for instance, \cite{ghose2019demystifying}).
	We conclude that \tech{} improves performance for multiple DDRx interface standards.

\subsection{Effect of different design thresholds}
\label{sec:design_threshold}
\subsubsection{RAPL limit:}
\begin{figure}[h!]
	\centering
	\centerline{\includegraphics[width=0.99\columnwidth]{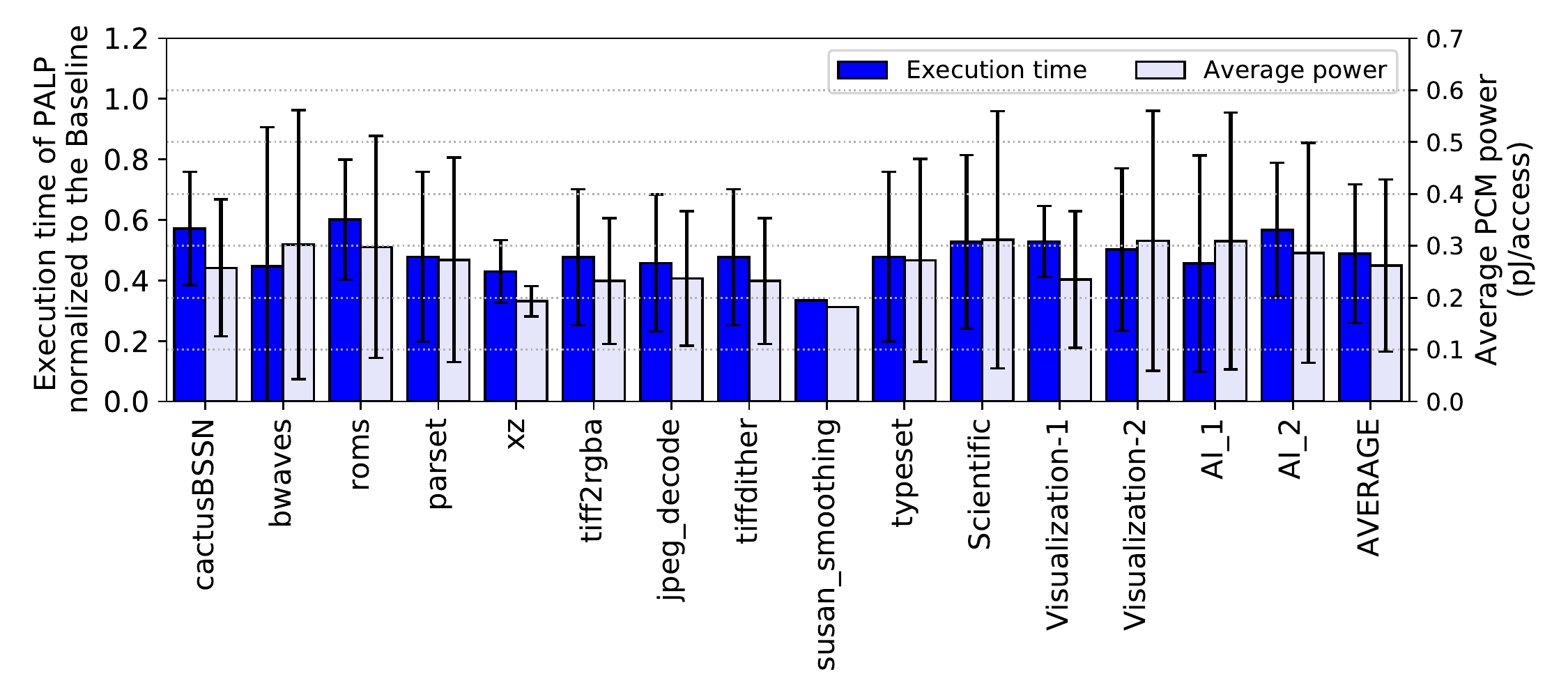}}
	\vspace{-10pt}
	\caption{Execution time with \tech{} normalized to the Baseline and average power consumption with \tech{} for all workloads. Height of each bar represents \tech{}'s result with the default RAPL limit of 0.3pJ/access \cite{lung2016double}. Error bars represent minimum to maximum variation obtained by sweeping the RAPL limit from 0.2pJ/access to 0.4pJ/access.}
	\label{fig:rapl_variation}
\end{figure}

Figure \ref{fig:rapl_variation} summarizes the variation in \tech{}'s execution time (normalized to the Baseline) and PCM power consumption for all workloads, when RAPL limit changes between 0.2-0.4 pJ/access. The bar heights represent the execution time and power consumption using the default RAPL limit of 0.3pJ/access. The error bars represent variation when the RAPL limit is varied between 0.2pJ/access and 0.4pJ/access. We make the following three observations.

	\textbf{First}, different RAPL limits lead to different performance and power consumption trade-offs in all workloads. In other words, the RAPL limit can be set to achieve the desired performance and power targets. 
	\textbf{Second}, setting a stricter RAPL limit results in lower performance (i.e., higher normalized execution time), while reducing the average power. We observe that for bwaves, setting the RAPL limit to 0.2 pJ/access results in a performance improvement of only 11\% over the Baseline, compared to the 33\% when RAPL limit is set to 0.4pJ/access.
	\textbf{Third}, beyond the RAPL limit of 0.4pJ/access, we see no significant variation in either execution time or power consumption, which means that the RAPL limit for PCM can be safely reduced from its rated value of 0.4pJ/access \cite{lung2016double}.

\subsubsection{Backlogging threshold $th_b$:}
\begin{figure}[h!]
	\centering
	\centerline{\includegraphics[width=0.99\columnwidth]{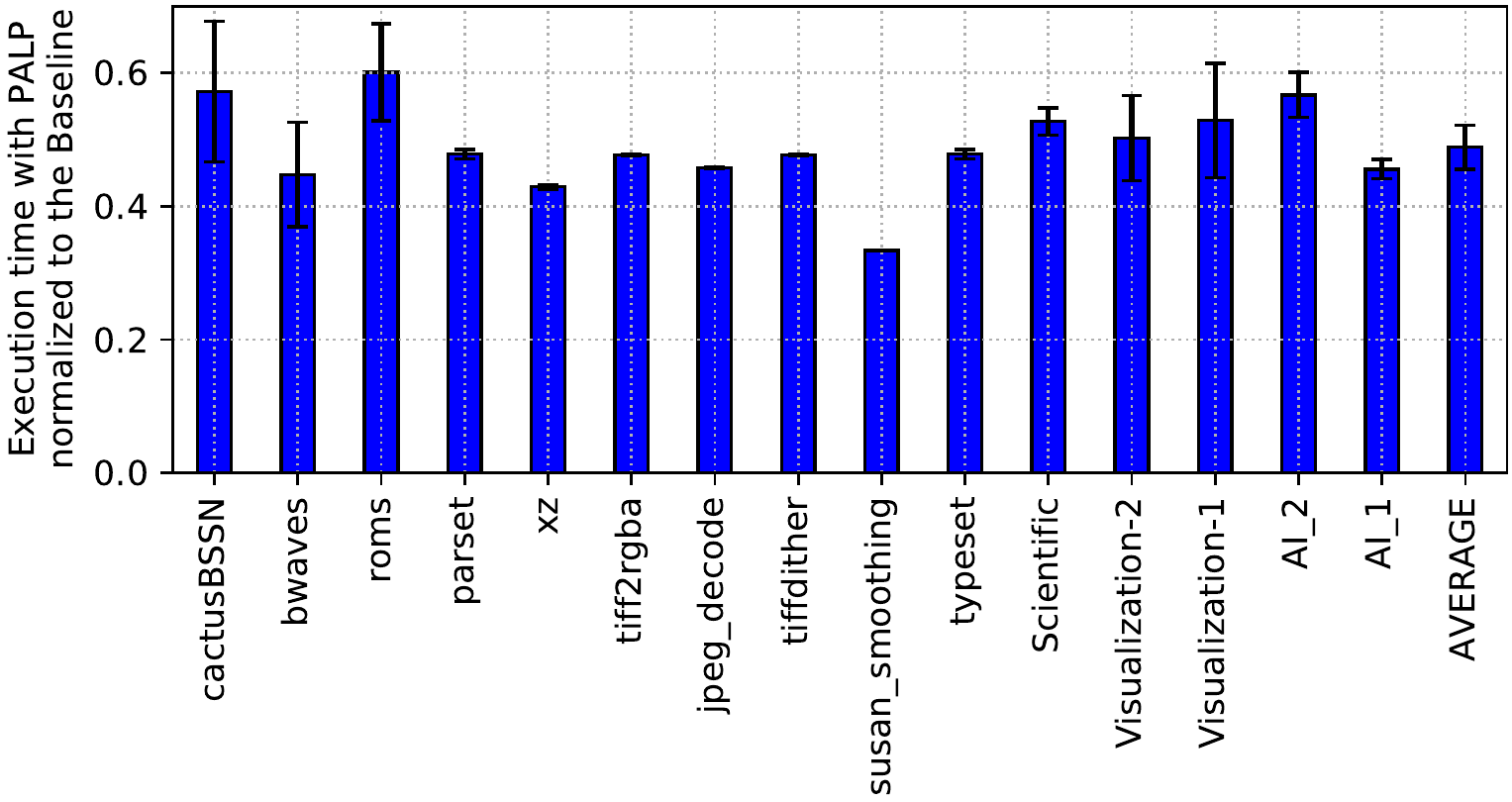}}
	\vspace{-10pt}
	\caption{Execution time with \tech{} normalized to the Baseline for all workloads. Height of each bar represents \tech{}'s execution time using the default backlogging threshold ($th_b$) of 8 accesses. Error bars represent minimum to maximum variation obtained by sweeping the threshold from 2 to 16.}
	\label{fig:bth_variation}
\end{figure}

\hl{Figure \ref{fig:bth_variation} summarizes the variation in \tech{}'s execution time normalized to Baseline for all workloads. Each bar height represents the execution time using the default backlogging threshold of 8 accesses. The error bars represent variation when the backlogging threshold is varied from 2 to 16. We make the following two observations.}

	\textbf{First}, normalized execution time changes as we vary the backlogging threshold. This is because setting a lower threshold forces \tech{} to schedule outstanding requests sooner, prioritizing starvation freedom over performance (i.e., parallelism exploitation). This leads to a reduction in \tech{}'s performance improvement. On the other hand, setting the backlogging threshold to a higher value offers more flexibility to \tech{}, allowing it to exploit partition-level parallelism more aggressively, thereby improving performance.	
	\textbf{Second}, for workloads such as xz, there is no impact of varying the backlogging threshold, meaning that \tech{} can prioritize starvation freedom for these workloads.



\subsection{Impact of \tech{} design decisions}

\subsubsection{Bank conflicts and scheduling:} 
\begin{figure}[h!]
	\centering
	\centerline{\includegraphics[width=0.99\columnwidth]{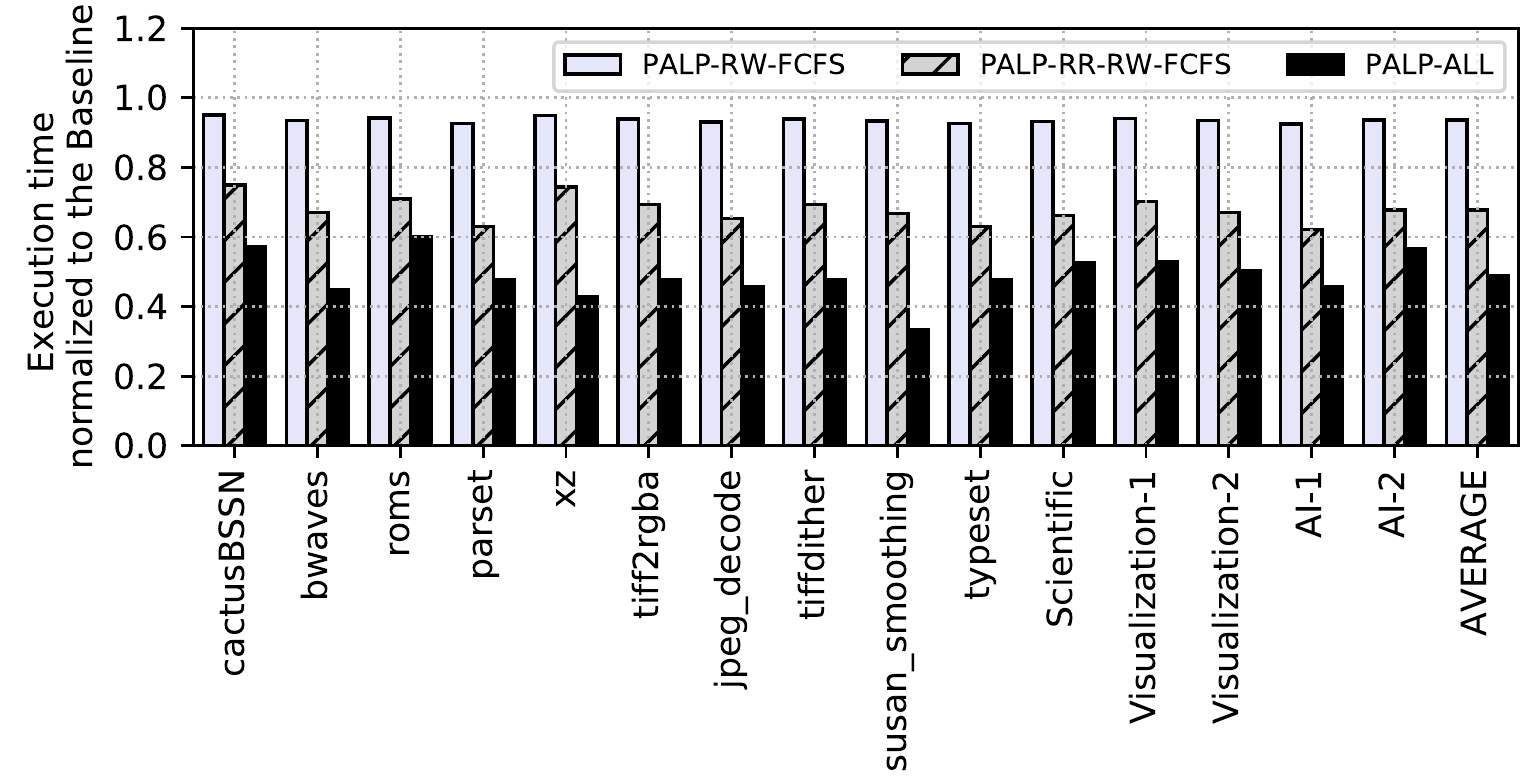}}
	\vspace{-10pt}
	\caption{Performance impact of \tech{}'s different components.}
	\label{fig:dc}
\end{figure}

\hl{To estimate the impact of resolving both the read-read and read-write bank conflicts, and the access scheduling policy, in Figure \ref{fig:dc} we report the execution time of \tech{} normalized to Baseline, with three configurations: (1) \tech{} resolving read-write conflicts only (\tech-RW-FCFS), (2) \tech{} resolving both conflicts (\tech-RR-RW-FCFS), and (3) \tech{} resolving both conflicts with the new access scheduling policy (\tech-ALL). We make the following two observations.
\textbf{First}, by resolving the read-write bank conflicts, \tech{} improves performance by only 7\% over the Baseline.\footnote{\tech{}-RW-FCFS is comparable to the original design proposed in the MultiPartition technique \cite{zhou2016efficient}. However, the results we present in Section \ref{sec:performance_improvement} for the MultiPartition technique are better than those we present here for \tech{}-RW-FCFS. This is because, in Section \ref{sec:performance_improvement}, we implemented out-of-order scheduling for the MultiPartition technique, to enable a more fair comparison with our \tech{} mechanism.
} When \tech{} resolves both read-write and read-read bank conflicts, performance improves by 32.2\% over the Baseline.
\textbf{Second}, by introducing our new scheduling policy, \tech{}'s performance improves significantly (51.1\% lower average execution time than the Baseline with \tech-ALL). We \textbf{conclude} that our choice of resolving both the read-read and read-write conflicts, and the new access scheduling policy are all \textbf{essential} to provide the highest performance benefits with \tech{}.}

%% file: sections/related_works.tex
To our knowledge, this is the \textit{first} work that 1) enables and exploits partition-level parallelism in phase-change memory to resolve read-read bank conflicts, and 2) designs a new memory access scheduling mechanism to aggressively exploit PCM's partition-level parallelism.


\subsection{Read-While-Write in PCM}
A  patent application \cite{barkley2017apparatus} describes \texttt{read-while-write} for PCM, where a read and write request can be scheduled simultaneously from a PCM bank using different partitions. However, no architectural technique is described on how to leverage this feature for system performance. Some earlier works such as \cite{zhou2016efficient} address architectural aspects assuming unrealistic system settings (such as infinite memory channel bandwidth). Our work not only addresses limitations of these prior works to resolve read-write bank conflicts, but also resolves read-read bank conflicts for the first time. We also evaluate \tech{} against  a realistic version of \cite{zhou2016efficient}  and find that \tech{} improves average system performance by \perf{}.

\subsection{Performance/energy/endurance improvement of PCM}
Many prior works optimize performance and energy of PCM \cite{LeeMicro10,yoon2015efficient,lee2010phase,QureshiISCA09,QureshiISCA12,ArjomandISCA16,cho2009flip}. Cho et al. propose Flip-N-Write \cite{cho2009flip} to improve PCM performance by first reading the memory content and then programming only the bits that need to be altered. Qureshi et al. propose PreSET \cite{QureshiISCA12}, an architectural technique that SETs the PCM cells of a memory location in the background before programming them during write. This improves performance by converting a write operation to a RESET operation of the PCM cells, which is faster. There are also techniques to consolidate multiple write operations \cite{xia2014dwc} to reduce the number of cells that need to be programmed, saving energy and improving performance. 
To mitigate PCM's cell-level endurance problem, several wear-leveling techniques are proposed \cite{SeongSecurityISCA2010,akram2018write}. 
\tech{} can be combined with these and similar techniques.


\subsection{Writeback optimization}
\mc{Several prior works propose line-level writeback \cite{QureshiISCA09,LeeMicro10,lee2010phase,LeeISCA2009,pourshirazi2018wall,pourshirazi2019writeback}, where for each evicted DRAM cache block, processor cache blocks that become dirty are tracked and selectively written back to PCM.
Various works propose dynamic write consolidation \cite{xia2014dwc,StuecheliISCA,wang2013wade,lee2010dram,SeshadriISCA14}, where PCM writes to the same row are consolidated into one write operation. 
Other works propose
write activity reduction \cite{hu2013write,huang2011register}, where registers are allocated on CPUs to reduce costly write operations in PCM.
Yet some other works propose multi-stage write operations \cite{yue2013accelerating,zhang2016mellow}, where a write request is served in several steps rather than in one-shot to improve performance.
Qureshi et al.  propose a morphable PCM system \cite{qureshi2010morphable}, which dynamically adapts between high-density and high-latency MLC PCM and low-density and low-latency single-level cell PCM.
Qureshi et al. propose write cancellation and pausing \cite{qureshi2010improving}, which allows PCM reads to be serviced faster by interrupting long PCM writes.
Jiang et al. propose write truncation \cite{jiang2012fpb}, where a write operation is truncated to allow read operations, compensating for the loss in data integrity with stronger ECC.
\tech~is complementary to all these approaches.
}

\subsection{Multilevel Cell PCM Optimizations}
PCM cells can be used to store multiple bits per cell (referred to as multilevel cell or MLC).
MLC PCM offers greater capacity per bit at the cost of asymmetric energy and latency in accessing the bits in a cell.
Yoon et al. propose an architectural technique for data placement in MLC PCM \cite{yoon2015efficient}, exploiting energy-latency asymmetries.
These techniques are also complementary to and can be combined with \tech{}.

%% file: sections/conclusions.tex
We introduce \tech{}, a new mechanism that enables each PCM bank's partition-level parallelism, and exploits such parallelism using a new memory access scheduling mechanism.
Previous architectural solutions to address parallelism in PCM banks can resolve only the read-write bank conflicts and assume an unrealistic memory interface with no timing constraints. We observe that (1) read-read bank conflicts far outnumber read-write bank conflicts, and (2) without designing a memory interface with realistic timing, the estimated performance improvements can be misleading. 
Based on our observations, we introduce \tech{}, which
is built on three contributions.
{First}, 
we introduce \emph{new} PCM commands to 
enable parallelism in a bank's partitions in order to 
resolve {read-write} bank conflicts, with \textit{modest} changes needed to PCM logic or its interface. 
{Second}, we propose \textit{simple} circuit modifications to resolve {read-read} bank conflicts.
{Third}, we propose a \emph{new} PCM access scheduling mechanism that improves performance by prioritizing those requests that exploit PCM bank's {partition-level parallelism} over other requests.
While doing so, our new scheduling mechanism also guarantees starvation-freedom and the running-average-power-limit (RAPL) of PCM.

We evaluate \tech{} with workloads from the MiBench and SPEC CPU2017 Benchmark suites.
Our results show that \tech{} reduces average PCM access latency by \alat{}, and improves average system performance by \perf{} compared to the state-of-the-art approaches. 

We \textbf{conclude} that \tech{} is a \emph{simple yet powerful} mechanism to improve PCM performance. We have open-sourced our infrastructure \cite{simulator} to enable future work based on \tech{} .